\documentclass{aa}  

\usepackage{graphicx}
\usepackage{graphicx}	
\usepackage{amsmath}	
\usepackage{subfigure}
\usepackage{subcaption}
\usepackage{hyperref}
\usepackage{natbib}
\usepackage[dvipsnames]{xcolor}
\usepackage[normalem]{ulem}
\usepackage{float}
\usepackage{booktabs}
\usepackage{txfonts}
\defcitealias{Pavan2025}{P25}
\defcitealias{Dreas2025}{D25}

\begin{document}

   \title{Evolution and afterglow emission of gamma-ray burst jets\\ from binary neutron star mergers}

   \author{E. Dreas
          \inst{1,2,3}\fnmsep\thanks{edreas@sissa.it}
          ,
          O. S. Salafia\inst{4,5},
          A. Pavan\inst{2,6},
          R. Ciolfi\inst{2,6},
          \and
          A. Celotti\inst{1,3,4,7}
          }

      \institute{SISSA, Via Bonomea 265, I-34136 Trieste, Italy            
        \and
             INAF, Osservatorio Astronomico di Padova, Vicolo dell'Osservatorio 5, I-35122 Padova, Italy
        \and
        INFN, Sezione di Trieste, Via Valerio 2, I-34127 Trieste, Italy
        \and
        INAF, Osservatorio Astronomico di Brera, via Bianchi 46, I-23807 Merate, Italy
        \and
        INFN, Sezione di Milano-Bicocca, Piazza della Scienza 3, I-20126 Milano, Italy
        \and
        INFN, Sezione di Padova, Via Francesco Marzolo 8, I-35131 Padova, Italy
        \and
        IFPU, Via Beirut 2, I-34151 Trieste, Italy
             }
   \date{Received xxx; accepted xxx}

 
  \abstract
   {Relativistic jets launched in binary neutron star (BNS) mergers are widely accepted as the engines powering most of the population of short gamma-ray bursts (GRBs). Understanding their structure and dynamics—particularly during and after breakout from the merger ejecta—is crucial for interpreting GRB afterglows, especially for off-axis observers. Traditional models often assume simple angular or radial jet profiles, potentially missing key features emerging for jets piercing through realistic environments.}
   {This work aims to investigate how the interaction between relativistic jets and the non-homogeneous, anisotropic environment of BNS mergers shapes the outflow structure, and how these features affect the resulting afterglow emission, across a range of jet configurations.}
   {We perform a series of 3D relativistic magnetohydrodynamic simulations of jets launched in post-merger environments, exploring different injection conditions. Simulations are evolved to late times, approaching the ballistic regime, where further dynamical evolution becomes negligible. From the resulting outflows, we extract energy and velocity profiles and compute multi-wavelength afterglow light curves using a semi-analytic model that includes radial stratification and the full 3D jet geometry.}
   {More energetic or earlier-launched jets drill more efficiently through the ejecta, but all develop asymmetries that leave clear imprints in the off-axis afterglow light curves. All models exhibit a complex multi-shock breakout structure responsible for an early, dimmer peak in the afterglow. Using the GW170817 afterglow as a reference case study, we show that all simulated jet structures can be brought into reasonable agreement with the observational data within the explored parameter space.}
   {}


   \titlerunning{Evolution and afterglow emission of GRB jets from BNS mergers}
\authorrunning{E.~Dreas et al.}
   \maketitle
%

\section{Introduction}
\label{sec:intro}

Gamma-ray bursts (GRBs) are the most energetic explosions in the universe, observed as brief flashes of gamma rays followed by longer-lasting broadband afterglow emission. They are believed to originate from two main types of progenitors: the collapse of massive stars and the mergers of compact binary systems involving neutron stars or black holes (\citealt{Eichler1989}; \citealt{Narayan1992}; \citealt{Woosley1993}; \citealt{Berger2014}).
The multi-messenger detection of GW170817 and its associated GRB 170817A prompted extensive multi-wavelength observations of a GRB originating from a binary neutron star merger (\citealt{LVC-BNS,LVC-MMA,LVC-GRB}), offering a detailed view of the afterglow evolution and confirming that these systems can launch highly relativistic, collimated outflows \citep{Mooley2018b,Ghirlanda2019}.
In this work, we focus on GRBs arising from binary neutron star (BNS) mergers. 

Numerical simulations -- especially those using relativistic magnetohydrodynamics (RMHD) -- can offer a realistic treatment of the jet launching and early evolution, incorporating magnetic fields, collimation shocks, and time-dependent energy injection from the central engine (e.g. \citealt{Komissarov2009}; \citealt{Tchekhovskoy2010}; \citealt{Gottlieb2023b}).
The afterglow emission is produced when the relativistic ejecta interact with the `circum-burst' medium surrounding the progenitor, generating shocks that accelerate electrons and amplify magnetic fields, resulting in synchrotron emission observable across the electromagnetic spectrum (\citealt{Meszaros1997}; \citealt{1998ApJ...497L..17S}). Modelling this afterglow emission is crucial for interpreting observed light curves and spectra, as well as for constraining key physical parameters such as the jet energy, circum-burst density, and angular structure of the outflow (see e.g. \citealt{Nakar2020}; \citealt{granot2014}).

While analytic and semi-analytic afterglow analyses have successfully captured many observational features, they often rely on idealised assumptions and simplified jet structures, which may not accurately reflect the physical properties of the jet material. 
Indeed, there remains a disconnection between the early-time evolution modelled in RMHD simulations and the late-time behaviour of the outflow, where the jet becomes kinetically dominated and asymptotically approaches a ballistic expansion regime. Bridging this gap requires robust methods for extracting the physical state of the jet from simulations and evolving it up to where radiative processes become observable.

Here, we present a framework to model the afterglow emission of GRB jets using the data extracted from RMHD simulations, with a particular focus on the transition to ballistic expansion, following the work presented in \citealp{Dreas2025}. Our approach integrates RMHD simulation outputs with semi-analytic hydrodynamical and radiative calculations, enabling a consistent mapping from the simulation domain to synthetic afterglow observables. We aim to quantify how the features imprinted during the jet propagation, breakout and initial expansion phase affect the observed light curves and viewing angle dependence. To this end, we exploit the set of jet simulations presented in \citealp{Pavan2025} (hereafter \citetalias{Pavan2025}), which explore variations in the injection parameters around a reference configuration. 
The main contribution of this work is the direct computation of afterglow observables from three-dimensional jet structures shaped by a realistic merger environment, using GW170817 as a well-observed reference case to anchor the model. Within this framework, we compare the jet structures and the resulting light curves obtained for the different injection conditions explored in \citetalias{Pavan2025}.

\begin{table*}
\caption{Jet injection parameters and main properties at $t_\mathrm{end}$ for all jet models.}
\label{tab:models}
\centering
\begin{tabular}{lccccccccc}
\hline\hline
Model & \multicolumn{4}{c}{Injection parameters} & \multicolumn{5}{c}{Properties at $t_\mathrm{end}=9.5$\,s} \\
\cmidrule(lr){2-5} \cmidrule(lr){6-10}
 & $t_\mathrm{j}$\,[ms] & $L_\mathrm{j}$\,[erg\,s$^{-1}$] & $\tau_\mathrm{d}$\,[ms] & $\Sigma_\mathrm{j}$ & $E_\mathrm{iso,max}$\,[erg] & $\overline{\Gamma}_\mathrm{max}|_\mathrm{\Phi_\mathrm{avg}}$ & $E_\mathrm{kin}/E_\mathrm{sum}$ & $\Theta_{50}\,[^\circ]$ & $\Theta_\mathrm{off}\,[^\circ]$ \\
\hline
\textit{F} & $385$ & $5.37\times10^{51}$ & $300$ & $1.20\%$ & $1.46\times10^{53}$ & $27.7$ & $0.97$ & $5.6$ & $1.4$ \\
\textit{a}\ ($L_\mathrm{j}\uparrow$) & $385$ & $8.95\times10^{51}$ & $300$ & $1.20\%$ & $4.57\times10^{53}$ & $40.7$ & $0.97$ & $5.1$ & $1.6$ \\
\textit{b}\ ($\tau_\mathrm{d}\uparrow$) & $385$ & $5.37\times10^{51}$ & $500$ & $1.20\%$ & $3.93\times10^{53}$ & $80.1$ & $0.95$ & $4.9$ & $1$ \\
\textit{c}\ ($B_\mathrm{j}\uparrow$) & $385$ & $5.37\times10^{51}$ & $300$ & $1.56\%$ & $1.94\times10^{53}$ & $26.8$ & $0.97$ & $5.0$ & $2.7$ \\
\textit{d}\ ($B_\mathrm{j}\uparrow, L_\mathrm{j}\downarrow$) & $385$ & $4.12\times10^{51}$ & $300$ & $1.56\%$ & $1.26\times10^{53}$ & $34.2$ & $0.96$ & $5.2$ & $1.5$ \\
\textit{e1}\ ($B_\mathrm{j}\downarrow, L_\mathrm{j}\downarrow$) & $385$ & $3.69\times10^{51}$ & $300$ & $0.46\%$ & $0.97\times10^{53}$ & $23.6$ & $0.96$ & $6.1$ & $3.6$ \\
\textit{e2}\ ($B_\mathrm{j}\downarrow, L_\mathrm{j}\downarrow, t_\mathrm{j}\downarrow$) & $185$ & $3.69\times10^{51}$ & $300$ & $0.46\%$ & $2.33\times10^{53}$ & $54.1$ & $0.97$ & $4.6$ & $0.9$ \\
\hline
\end{tabular}
\tablefoot{The injection parameters considered in the study are the jet launching time (with respect to merger) $t_\mathrm{j}$, injected luminosity $L_\mathrm{j}$, decay time scale of the injected luminosity $\tau_\mathrm{d}$, and relative contribution of the magnetic field to the total luminosity $\Sigma_\mathrm{j}$. The table also shows, for each model, some properties of the jet cap (see text for definitions) at the end of the extended simulation, namely the maximum isotropic equivalent energy $E_\mathrm{iso,max}$, the maximum of the radially- and azimuthally-averaged Lorentz factor $\overline{\Gamma}_\mathrm{max}|_\mathrm{\Phi_\mathrm{avg}}$, the ratio of kinetic over total (excluding rest mass) energy $E_\mathrm{kin}/E_\mathrm{sum}$, the effective half-opening angle $\Theta_{50}$, and the offset of the jet barycentre with respect to the injection axis $\Theta_\mathrm{off}$. The parameters shown next to each model label denote deviations (increases or decreases) from the fiducial model \textit{F}.}
\end{table*}

This work builds on recent efforts to incorporate simulation-informed jet structures into afterglow models (e.g. \citealt{gottlieb2018}; \citealt{Lazzati2018};  \citealt{Gill2019}; \citealt{Kathirgamaraju2019}; \citealt{Nativi2021, Nativi2022}; \citealt{Gottlieb2022a}). 
Our approach  considers the evolution of relativistic jets within a realistic, non-homogeneous BNS merger environment (\citealt{Ciolfi2020a}, \citealt{Pavan2021,Pavan2023}, \citealt{Pais2024}), capturing the complexity of the resulting jet-ambient medium interaction.

By continuing the simulation into the ballistic regime, we capture the full transition from jet breakout to free expansion, ensuring that the jet structure used for afterglow predictions reflects its final dynamical state. The resulting 3D energy and velocity distributions are then used as input to compute afterglow emission in the framework of the external shock model, where the decelerating jet interacts with the interstellar medium.

In this context, the radial stratification of the outflow -- i.e., the distribution of energy across shells with different Lorentz factors -- plays a critical role in shaping the light curve (e.g.\ \citealt{Panaitescu1998,Sari2000,Zhang2006, Lamb2020, Fraija2022, Moss2023}). As slower-moving shells catch up with the decelerating shock front, they inject additional energy into the blast wave, modifying the afterglow dynamics. This energy injection process affects both the timing and morphology of the afterglow rise, and can introduce characteristic features such as shallow rises or plateaus, especially for off-axis observers.

The paper is structured as follows. In Section~\ref{sec:jet_params}, we present the simulation setup, the various jet models and the different results in terms of dynamical evolution and energetics. In Section~\ref{sec:angular}, we analyse the angular distribution of the jet material and compare the different resulting structures. In Section~\ref{sec:afterglow}, we use such jet structures to compute the afterglow light curves with an enhanced afterglow model. Finally, in Section~\ref{sec:conclusion} we discuss our results.

\section{Impact of injection parameters on jet evolution}
\label{sec:jet_params}

\citetalias{Pavan2025} investigated how varying injection parameters -- namely the jet’s initial luminosity, decay time scale, magnetisation, and launching time -- affect the evolution of a jet propagating through a realistic post-merger environment, i.e.~directly imported from a BNS merger simulation \citep{Ciolfi2020a}. Here, we summarise the key outcomes during the first three seconds of evolution, as reported in \citetalias{Pavan2025}, and extend the analysis by following the jet dynamics up to nine seconds.
We note that the seven models considered (Table \ref{tab:models}) sample a limited region of the injection parameter space around a fiducial configuration. We further remark that all the simulations presented in this work are performed at a fixed numerical resolution, and are not in a numerical convergence regime. The results discussed in the following are therefore indicative of the behaviour of each configuration: we expect the trends to be preserved, while the specific values may differ in a converged regime. A discussion of the dependence of the jet properties on resolution is presented in Appendix~\ref{app:resolution}.

\subsection{Numerical setup}
\label{sec:setup} 

All simulations are performed using the RMHD module of the PLUTO code \citep{Mignone2007-PLUTO1}. 
The first three seconds of evolution following the jet launching time, $t_\mathrm{j}$, are reproduced on a 3D spherical computational grid, as detailed in \citetalias{Pavan2025}.
Then, at $ t-t_\mathrm{j}=3$\,s, the simulation is transferred to a Cartesian grid using the remapping procedure outlined in \citealt{Dreas2025} (hereafter D25), which enables the evolution to be followed up to $ t_\mathrm{end} \equiv t-t_\mathrm{j}=9.5$\,s without loss of resolution. 
In particular, we focus on the evolution of the northern side of the jet, as in D25, though the same procedure can be equally applied to the southern side.

Specifically, the Cartesian grid spans $x, z \in [-2.33, +2.33] \times 10^6$ km and $y \in [0.3, 6.3] \times 10^6$ km, with the $y$-axis being aligned to the main jet propagation direction. The grid resolution is uniform along all axes, with $656\times864\times656$ points along the $x$-, $y$- and $z$-axes, respectively.
Numerical integration is carried out employing piecewise parabolic reconstruction, the HLL Riemann solver, third-order Runge–Kutta time stepping, and hyperbolic divergence cleaning to enforce $\nabla \cdot \mathbf{B} = 0$ \citep[][]{MT2010,MTB2010}. Furthermore, a Taub-Matthews equation of state (TM EOS hereafter) is adopted to relate thermodynamic quantities (\citealt{Mignone2005}, \citealt{Mignone2007}). 

Below, we discuss the results of our simulations for the various jet models considered, whose injection parameters and properties at $t_\mathrm{end}$ are summarised in Table~\ref{tab:models}.

\subsection{Fiducial model}
\label{sec:fiducial}

Our fiducial model \textit{F}, used as the baseline for varying injection parameters in the other models, describes the evolution of a manually injected jet at $t_\mathrm{j}=385$\,ms after merger, with an initial luminosity of $L_\mathrm{j}=5.37\times10^{51}$\,erg/s, exponentially decaying with a time scale of $\tau_\mathrm{d}=300$\,ms, and magnetic field contribution to the total injected luminosity of $\Sigma_\mathrm{j}=1.2\%$ (see Section~2.1 in \citetalias{Pavan2025}). 

\begin{figure*}
    \centering
    \includegraphics[width=0.9\linewidth]{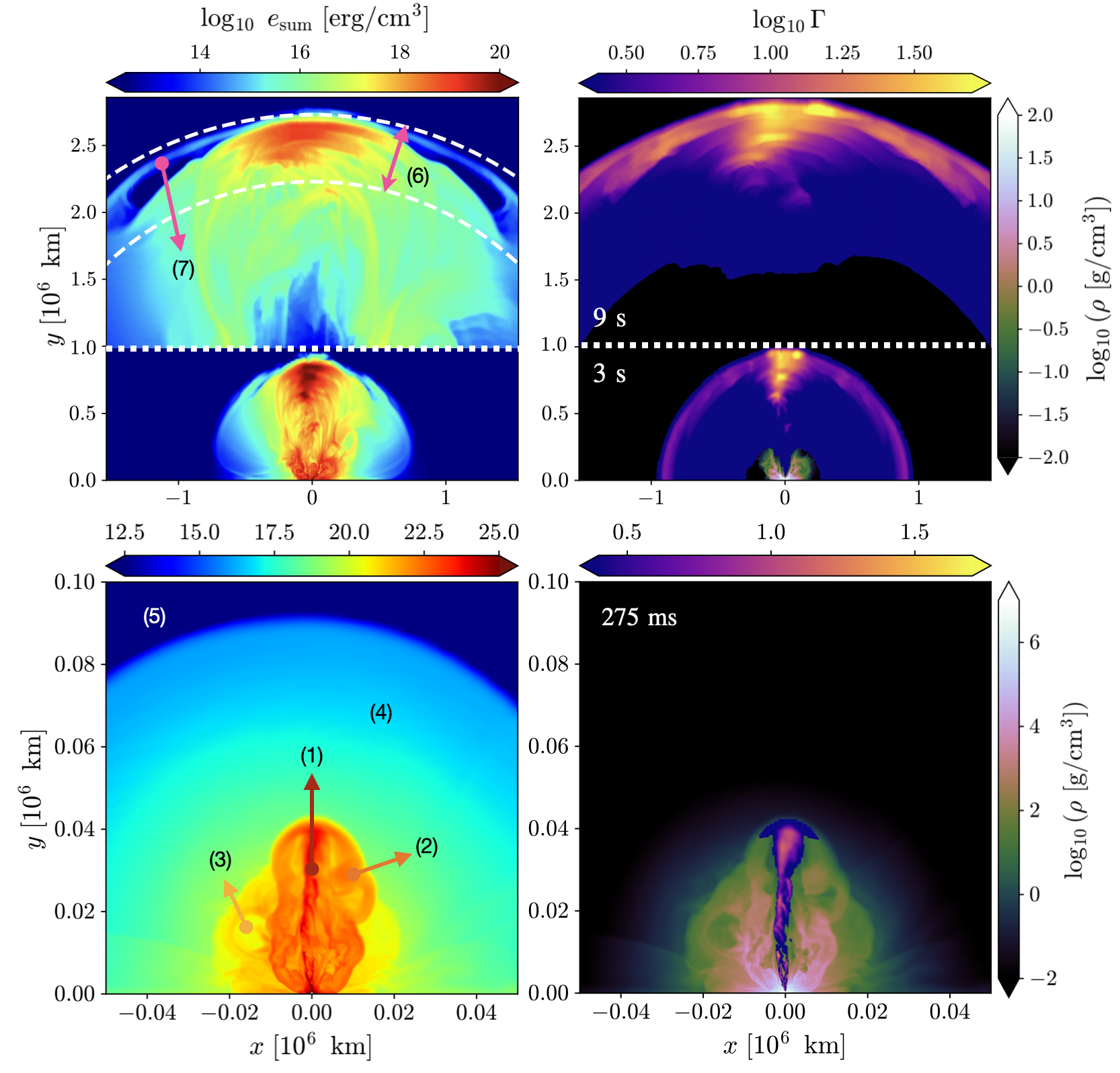}
    \caption{Meridional view of total energy density excluding rest-mass contribution ($e_\mathrm{sum}$; left column) and rest-mass density and Lorentz factor (right column), at $275$\,ms (bottom), $3$\,s, and $9$\,s (top) after launch in the fiducial model \textit{F}. The threshold in $\Gamma$ is 1.5, 1.05 and 1.4 respectively for the three representative times. 
    The top row shows composite panels combining snapshots at $3$ and $9$\,s, referring to the former up to $y\!=\!10^6$\,km and to the latter above it.
    See text for the definition of the regions identified by numbers.}
    \label{fig:shell_evolution}
\end{figure*}

Fig.~\ref{fig:shell_evolution} shows the configuration of the jet-environment system for the fiducial model, in the meridional plane at three key stages: at 275\,ms after launch (lower panel), at the end of the first three seconds of evolution within the spherical grid, and at $t_\mathrm{end}$ (upper panels). 

In this model, the jet's evolution is characterised by intense mixing with the realistic BNS merger environment. At $t_\mathrm{j}\approx275$\,ms the jet and cocoon, regions (1) and (2) in Fig.~\ref{fig:shell_evolution}, pierce through the inner merger environment, region (3). We refer to this moment as the breakout time (see the lower panels in the Figure). Before the breakout, approximately half of the injected energy is deposited into the inner environment (see Table~2 in \citetalias{Pavan2025}).
Following breakout, the jet head undergoes a sharp transition as it emerges from the dense environment and begins accelerating into a much lower-density medium.
The distribution of the corresponding energy density, without the rest mass component ($e_\mathrm{sum}$, left column), reveals an isotropic bubble of non-negligible energy, corresponding to the dynamical ejecta imported from the merger simulation. We refer to this region as the outer merger environment (indicated in the Figure as region 4).
At approximately 3\,s after launch, the jet head reaches the boundary of the outer merger environment, having accelerated up to a maximum Lorentz factor of $\simeq56$.

The jet ejecta propagation within the atmosphere (region 5), after $t-t_\mathrm{j} = 3\,$s, is followed in the uniform Cartesian grid (see Section~\ref{sec:setup}).
As the evolution proceeds, the system enters a phase of homologous expansion, during which the shocked material settles into a multi-component, stratified structure. Specifically, we identify the \textit{jet cap}, region (6) in the Figure, as the zone containing most of the jet energy. The jet cap is formally defined as the radial extent where the kinetic energy flux is above 15\% of its maximum\footnote{See \citetalias{Pavan2025} and refs. therein.}. Note that in our previous works we referred to this region as the `jet head': here we update the terminology to avoid confusion with its canonical definition. The jet cap is preceded by a thin shell of shocked material, region (7), to which we will refer as \textit{outer shell} (see Section~\ref{sec:2shell} for further discussion). 

During this phase, the jet ejecta undergo residual conversion of the retained thermo-magnetic energy into kinetic form, along with the development of a stable angular structure (see further discussion in Section~\ref{sec:energy}).

\subsection{Influence of luminosity and decay time scale}
\label{sec:inj_energy}

Models \textit{a} and \textit{b} represent more energetic jets compared to the fiducial case. Specifically, model \textit{a} features an initial luminosity that is $\approx\!170\%$ of that of model \textit{F}, while model \textit{b} has the same jet luminosity as the fiducial case, but features a longer injection time scale $\tau_\mathrm{d}=500$\,ms, reaching the same total injected energy as model \textit{a} over the full simulation time span. 

As a result of its higher $L_\mathrm{j}$, model \textit{a} exhibits faster jet propagation than the other models, achieving breakout at $t_\mathrm{j}\approx200$\,ms.
Model \textit{b} follows a similar early-time evolution and breakout time as the fiducial model, but drills more efficiently through the same inner merger environment. 
Notably, both models \textit{a} and \textit{b} retain a larger fraction ($\approx\!60\%$) of injected energy at the jet cap compared to model \textit{F}.
This enhances jet acceleration into the outer merger environment, resulting in local maximum Lorentz factors of $85$ and $102$ in models \textit{a} and \textit{b}, respectively, at $t-t_\mathrm{j} = 3$\,s.

\subsection{Influence of magnetisation}
\label{sec:inj_mag}

Models \textit{c} and \textit{d} feature an increased magnetic field contribution ($\Sigma_\mathrm{j}=1.56\%$) with respect to model \textit{F}. 
Additionally, model \textit{d} employs a lower initial luminosity ($\approx\!75\%$ with respect to the fiducial) to assess the impact of magnetisation relative to jet power (see Table~\ref{tab:models}).

In both models, the higher initial jet magnetisation is associated with better preservation of the jet's toroidal magnetic field, owing to the suppression of hydrodynamical instabilities at the jet-environment interface (see Figures~6 and 7 of \citetalias{Pavan2025}).
This, in turn, results in higher jet collimation throughout propagation, despite the same (or even lower) initial jet luminosity relative to the fiducial case (further discussion in Section~\ref{sec:angular}).  
On the other hand, the jet dynamics and energetics remain similar between models \textit{c} and \textit{F}, whereas the reduced $L_\mathrm{j}$ in model \textit{d} leads to a less efficient jet drilling, resulting in a slightly longer breakout time ($t_\mathrm{j}\approx300$\,ms) and increased energy deposition into the surrounding environment (see \citetalias{Pavan2025}).

\subsection{Influence of launching time}
\label{sec:inj_time}

Finally, the impact of jet launching time is examined by considering two models with identical injection properties -- both with lower initial luminosity and magnetisation compared to the fiducial model, see Table \ref{tab:models} --  but different injection times, $t_\mathrm{j}\!=\!385$\,ms and $185$\,ms after merger for model \textit{e1} and \textit{e2}, respectively.

As expected, $t_\mathrm{j}$ has a significant impact on both the mass and radial extent of the merger environment that the jet needs to drill through. 
Specifically, in model \textit{e1}, the surrounding medium is approximately a factor 2 more massive and extends about 4 times farther at the time of jet launch than in model \textit{e2}.
For the low-luminosity, low-magnetisation model \textit{e1}, the jet undergoes severe disruption as it interacts with the merger environment, with more than half of the injected energy being deposited into the environment itself and a final jet angular structure that lacks a well-defined ultra-relativistic core (Section~\ref{sec:angular}). As this configuration lies close to the marginal, quasi-choked regime, some of its structural and observational properties are comparatively more sensitive to numerical resolution than those of the other models, as discussed in Appendix~\ref{app:resolution}.

In contrast, model \textit{e2} encounters a much less extended and less massive medium, reducing both the strength and duration of the jet–environment interaction. Notably, the jet breaks out in just $100$\,ms (the fastest among all models at the fiducial resolution), while retaining over half of the injected energy at the jet cap. In addition, the jet retains a collimated angular structure with an ultra-relativistic core and a toroidally dominated magnetic field configuration (see Figure~8 of \citetalias{Pavan2025}). 
These results highlight that the jet launching time plays a major role in determining the outcome of propagation in realistic BNS merger environments (see, e.g., \citealt{Murguia-Berthier2014}, \citealt{Lyutikov2020} and \citealt{Pais2024} for previous analyses on the jet choking conditions).

\subsection{Energy budget and ballistic regime}
\label{sec:energy}

In this Section, we summarise the evolution of the energy budget across the various models and describe the onset of the ballistic (or coasting) regime, which marks the point where the jet can be considered dynamically stable for afterglow modelling. 
Note that, unless stated otherwise, the values of energy discussed in this Section are referred to the whole system, intended as jet, cocoon, inner and outer merger environment.

Model \textit{a}, with its increased $L_\mathrm{j}$, achieves the highest values of kinetic, thermal, and magnetic energy as soon as $\simeq150$\,ms after launch (we refer to the left panel of Figure~3 in \citetalias{Pavan2025}). 
Model~\textit{b} initially evolves similarly to the fiducial case, but due to its longer $\tau_\mathrm{d}$,  
ultimately reaches an energy budget comparable to that of model~\textit{a}.
By contrast, although models \textit{c} and \textit{d} feature increased magnetisation, model \textit{c} evolves similarly to the fiducial case, whereas model \textit{d}, with its lower $L_\mathrm{j}$, produces an overall less energetic system. 
Finally, the low-luminosity and low-magnetisation models \textit{e1} and \textit{e2}, which differ in launching time, yield very different outcomes: in the former, the strong jet-environment interaction leads to an overall lower-energy system, especially at the jet cap; in the latter, the much lower energy deposition into the environment results in a jet cap energetically comparable to those in models~\textit{a} and \textit{b}. 
To quantify the energy budget of the different models, in Table~\ref{tab:models} we list the maximum values of the isotropic equivalent energy ($E_\mathrm{iso,max}$; see Eq.~35 in \citetalias{Pavan2025}), computed at the jet cap at $t_\mathrm{end}$. 
We notice that models \textit{a} and \textit{b} yield comparable $E_\mathrm{iso,max}$, a factor $\simeq1.7-2$ larger than that of model \textit{e2}, and more than twice those of models \textit{F}, \textit{c}, and \textit{d}. Model \textit{e1} stands out, with a maximum energy being more than four times lower than those of models \textit{a} and \textit{b}.

During the evolution, thermal and magnetic energies are progressively converted into kinetic form, driving continued jet acceleration.
This process eventually saturates as the jet enters the ballistic regime. 
Specifically, at $t-t_\mathrm{j} = 3$\,s, the fraction of energy at the jet cap in kinetic form ranges from $86\%$ to $92\%$ across the models (refer to Table~2 in \citetalias{Pavan2025}).
At $t-t_\mathrm{j} = 6$\,s, such a fraction reaches approximately 95\% for model~\textit{a}, 93\% for model~\textit{b}, 96\% for model~\textit{F}, 94\% for models~\textit{c, d, e1}, and 96\% for model~\textit{e2}. 
Finally, by $t_\mathrm{end}$, all models approach a saturation regime, with kinetic energy fractions between $95\%$ and $97\%$.
Notably, the relative increase in kinetic energy between 3 and 6\,s is much larger than that between 6 and 9\,s, reflecting the saturating trend in residual jet acceleration.

\begin{figure}
    \centering
    \includegraphics[width=\linewidth]{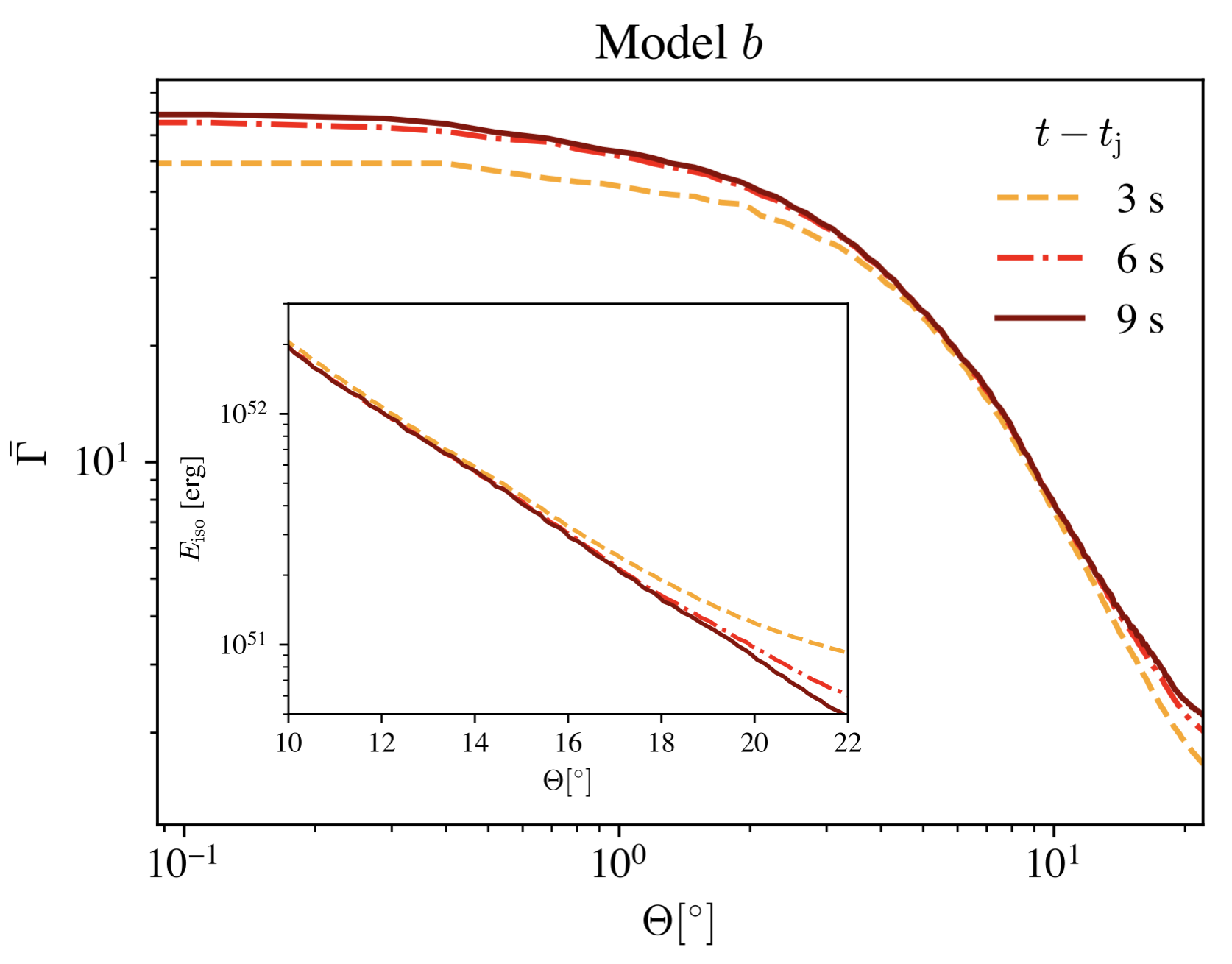}
    \caption{Radially and azimuthally averaged Lorentz factor, and the corresponding $E_\mathrm{iso}$ at $\Theta>10^{\circ}$ (inset), at the jet cap in model \textit{b} at different evolutionary times. See text for discussion.}
    \label{fig:1dgamma_evolution}
\end{figure}
\begin{figure}
    \centering
    \includegraphics[width=1\linewidth]{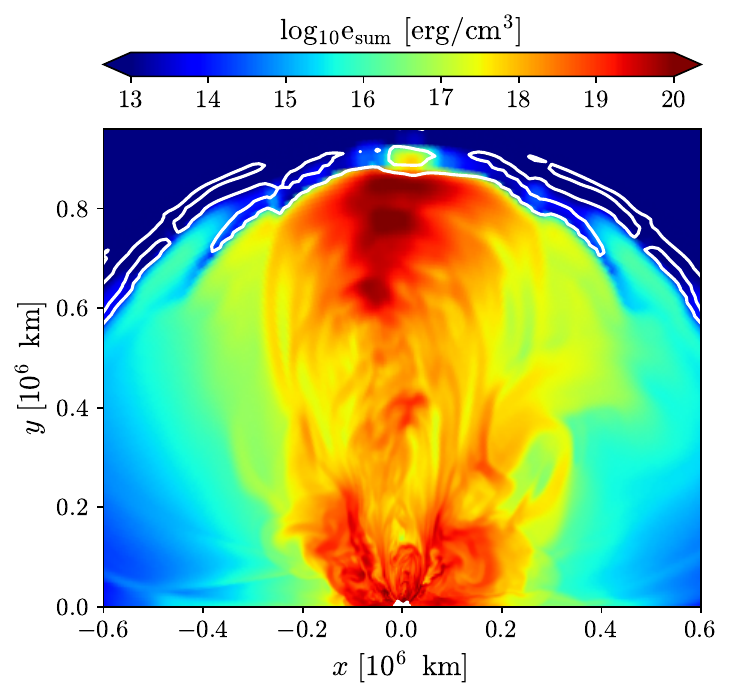}
    \caption{Meridional view of the total (minus rest-mass) energy density at $t-t_\mathrm{j}=3$\,s for model $F$. White contours identify shocked surfaces (see text).}
    \label{fig:shock_esum}
\end{figure}

The evolution of the bulk Lorentz factor at the jet cap offers a complementary view of the above acceleration process and its saturating trend. 
Tracking the maximum of the radially and azimuthally averaged (axisymmetrised) Lorentz factor ($\overline{\Gamma}_\mathrm{max}|_\mathrm{\Phi_\mathrm{avg}}$; see Eqs.~2 and 4 in D25), we find an average increase of about $4\%$ across the models, with the exception of model \textit{b}, which, owing to its longer $\tau_\mathrm{d}$, undergoes the most prolonged energy conversion period. As shown in Figure~\ref{fig:1dgamma_evolution}, the maximum relative change in $\overline{\Gamma}_\mathrm{max}|_\mathrm{\Phi_\mathrm{avg}}$ between $t-t_\mathrm{j} = 3$ and $6$\,s for model \textit{b} is approximately 34\%, decreasing to 14\% between 6 and $9$\,s ($t_\mathrm{end}$).

In the inset of Figure~2, we show the azimuthally averaged distribution of $E_\mathrm{iso}$ at $\Theta>10^{\circ}$, in model \textit{b}, which highlights the residual energy redistribution at large polar angles over time. Notably, the change in the slope of the distribution diminishes toward $t_\mathrm{end}$, confirming that the jet is approaching a stable angular configuration.

We therefore consider $t_\mathrm{end}$ a late enough  time to extract energy and velocity distributions for synthetic afterglow modelling: as shown above, the kinetic energy fraction is already close to unity for all models by $t_{\rm end}$, and its rate of increase, like that of $\overline{\Gamma}_{\rm max}|_{\Phi_{\rm avg}}$, has visibly slowed down between successive extraction times, indicating a saturating trend.

\subsection{Multiple shock structure}
\label{sec:2shell}

As described in Section \ref{sec:fiducial}, during its dynamical evolution the jet experiences a sharp transition as it emerges from the inner merger environment into a region of significantly lower external density, the outer merger environment, at the moment of breakout. The sudden change in density leads to the formation of a complex, multi-layered shock structure. 
At the forefront, a quasi-spherical outer shell propagates into the surrounding medium with an almost isotropic morphology, carrying a non-negligible energy budget compared to that of the cocoon itself. Notably, this leading shell is causally disconnected from the jet and its cocoon, which follow behind. Likely, its origin lies in the compression and acceleration of the fluid during the jet-cocoon breakout.
Figure \ref{fig:shock_esum} shows again the snapshot at $t-t_\mathrm{j}=3$\,s of $e_\mathrm{sum}$\footnote{For clarity, we report the plot only for a representative model (case $F$), as a similar structure develops for all models.}, where shocked regions are identified by the white contours as the boundaries where the fluid switches from subsonic to supersonic regime, i.e.~the Mach number transitions from less to greater than one. The sound speed for the TM EOS is computed using  Eq.~(17) of \cite{Mignone2007}.

The outer shell structure is of particular interest from an observational point of view, and is observed in all our jet simulations. At viewing angles $\theta_\mathrm{v} \gtrsim 15 ^\circ$, where the observer does not have a direct line of sight into the jet core, this shell contributes to an excess of energy at relativistic velocities in the angular distribution of the ejecta. In synthetic afterglow light curves, this manifests as an early, shallow peak in the emission, preceding the steeper rise typically associated with the decelerating cocoon or the structured jet core (see the discussion in Section \ref{sec:af_model}). This feature provides a distinct observational signature of the complex energy structure that the shock breakout imprints on the outflow.

\section{Angular distribution of the jet material}
\label{sec:angular}

Understanding the angular distribution of energy is crucial for interpreting afterglow light curves and for connecting theoretical models to observed GRB signatures, especially in cases of off-axis viewing angles. The morphology of the jet, its degree of symmetry, and the presence of fast-moving material at large angles all influence the observables.

\begin{figure*} 
   \centering
    \includegraphics[width=0.88\hsize]{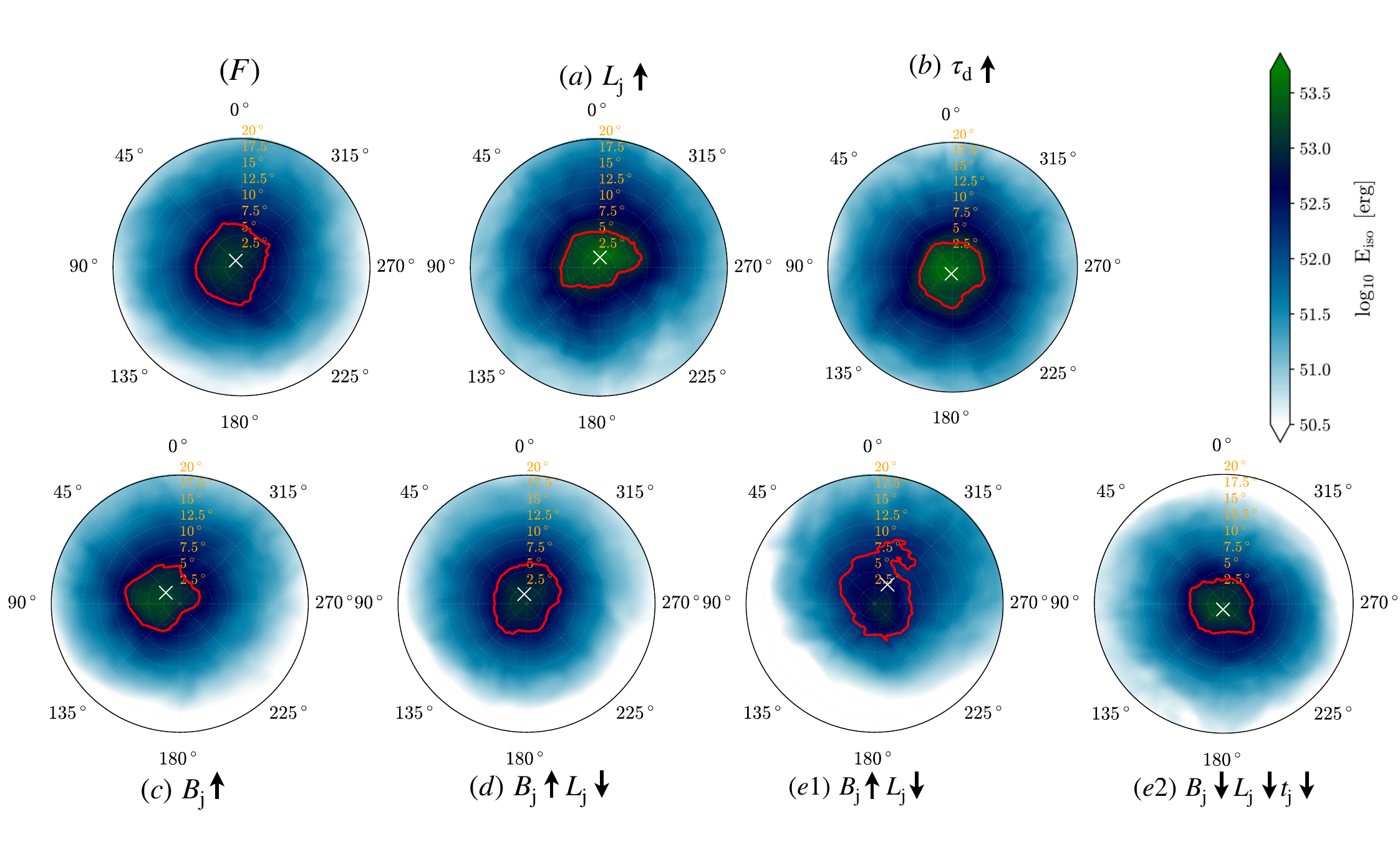}
      \caption{Front-view distributions of the isotropic-equivalent energy of the jet cap for all jet models at $t_\mathrm{end}$, with model names indicated in parentheses in each panel. The red line shows the smallest contour that contains 50\% of the energy; the white cross indicates the jet's energy barycentre.
              }
         \label{fig:frontview_en}
\end{figure*}

Figure \ref{fig:frontview_en} displays the front-view distribution of the isotropic-equivalent energy, $E_{\mathrm{iso}}$ (see Eq. 3 of \citetalias{Dreas2025}), evaluated in the jet cap at $t_\mathrm{end}$. 
The red line shows the smallest contour that encloses 50\% of the total energy, providing an estimate for the angular size of the jet’s energetic core. 
A white cross marks the jet’s barycentre, defined as the energy-weighted centroid in angular coordinates. The offset of this point from the centre ($\Theta_\mathrm{off}$) quantifies the jet’s axis displacement relative to the injection axis and provides insight into the degree of asymmetry introduced by the interaction with the environment or internal jet instabilities.
\begin{figure*}
    \centering
    \includegraphics[width=0.9\linewidth]{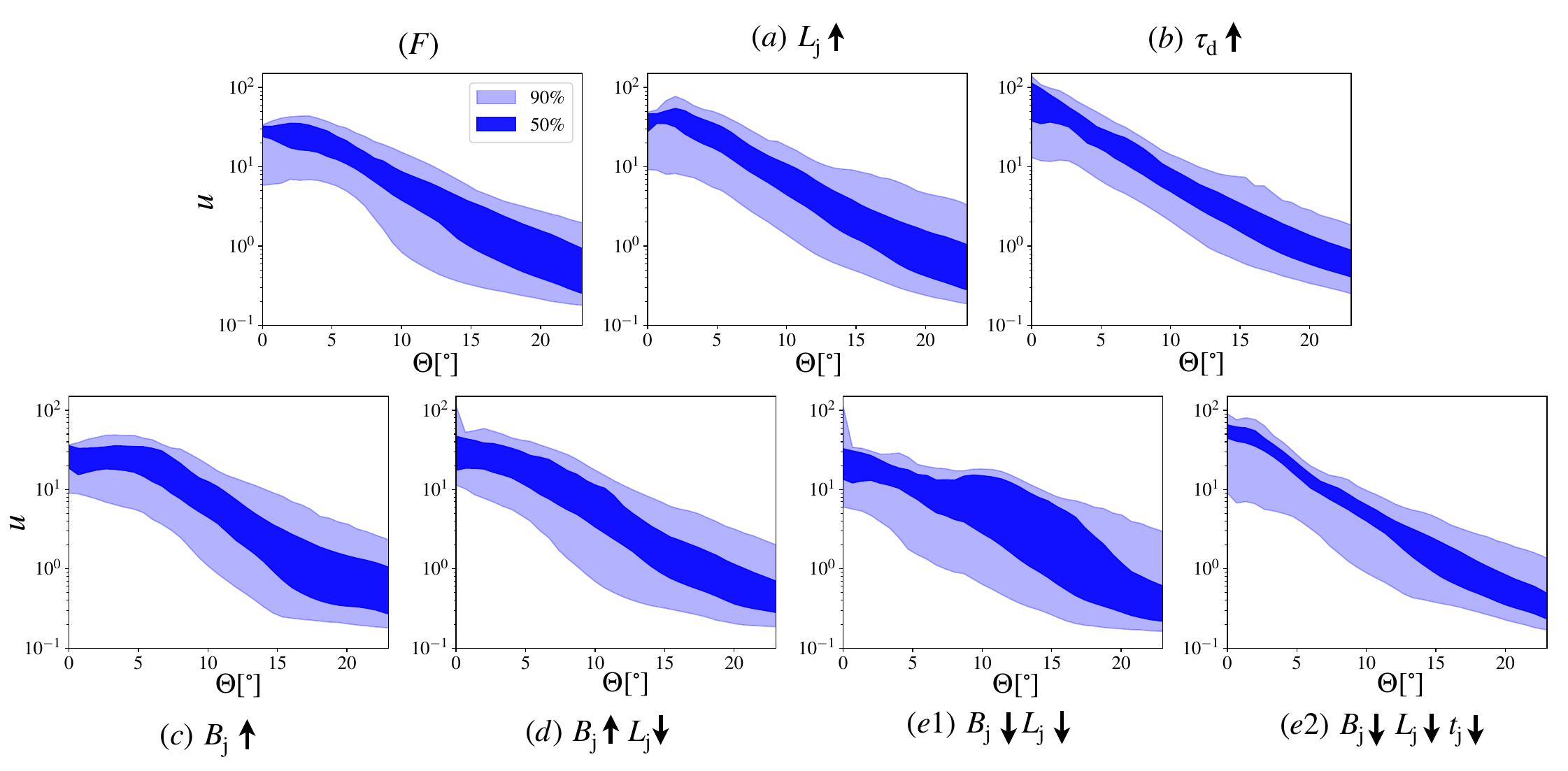}
    \caption{Jet energy distribution in the four velocity-polar angle plane. The dark blue shaded region identifies the interquartile range (25th–75th percentiles) while the range between the 5th and 95th percentiles is shown in light blue. Model names are indicated in parentheses in each panel} 
    \label{fig:vel_theta}
\end{figure*}
   
Indeed, none of the simulated jets shows an axisymmetric distribution, and we find that the barycentre is displaced from the injection axis by $0.9^\circ-3.6^\circ$ (see Table \ref{tab:models}). 

We define the effective half-opening angle $\Theta_{50}$ as the aperture angle of a cone subtending the same solid angle as that contained within the red contour. For all models, this is around $5^\circ$, with the smallest aperture given by model \textit{e2} ($4.6^\circ$), and the largest by model \textit{e1} ($6.1^\circ$, see Table \ref{tab:models}).
Models \textit{a} and \textit{b}, which have the highest injected energies, also feature the highest peak isotropic-equivalent energies. Model \textit{b} is particularly well-collimated and shows a roughly circular, centrally peaked energy profile, consistent with standard expectations for successful, structured GRB jets.

The impact of magnetisation is explored through models \textit{c} and \textit{d}, which exhibit only minor deviations from the fiducial case (\textit{F}) in terms of energy distribution and angular width. 

The comparison between models \textit{e1} and \textit{e2} -- which differ only in their jet launching time -- highlights the importance of the central engine’s timing relative to the collapse dynamics. Both models correspond to relatively weak jets, but their final angular structure differs significantly.  Model \textit{e2}, like model \textit{b}, develops a narrow, energetic, and well-centred core, while model \textit{e1} exhibits a less energetic, irregular core that is significantly misaligned. This underscores how earlier jet launching leads to reduced interaction with the ambient medium, preserving the jet’s structure.

To further characterise the jet, we examine the joint distribution of energy in angular and velocity space. Figure \ref{fig:vel_theta} shows the energy distribution in the $\Theta$–$u$ plane, where $u\!\equiv\!\Gamma\beta$ is the modulus of the four-velocity of the fluid in units of the speed of light. For each polar angle $\Theta$, we extract the percentiles of the energy distribution in $u$, calculated from the cumulative azimuthally averaged velocity profiles (see Appendix \ref{sec:velprof}). Specifically, we show the interquartile range (25th–75th percentiles, dark blue) and the broader range between the 5th and 95th percentiles (light blue). In the following, we refer to the n$^\mathrm{th}$ percentile for the four-velocity as $u_\mathrm{n}$. This diagnostic reveals not just how fast the fluid is moving at each angle, but how widely the energy is distributed across different speeds.

At small angles ($\Theta\!\lesssim\!4^\circ$), many profiles are relatively flat or display modest peaks around $\Theta\!\approx\!2^\circ$, which are likely a by-product of the jet axis misalignment. In these regions, the typical $u$ values are between 30 and 60, corresponding to ultra-relativistic bulk Lorentz factors. 
However, the substantial width of the distribution -- even at small angles -- reveals a broader velocity structure than commonly assumed. The presence of slower, though still relativistic material ($\Gamma\!\sim\!$ 3–10) within the core implies that the outflow is not a uniform shell, but a stratified structure composed of multiple shells with a relatively wide range of Lorentz factors.

Beyond $\Theta\!\approx\!4^\circ$, the median and upper percentile curves decline gradually, with $u_{95}$ remaining above 10 up to $\Theta\!\approx\!15^\circ$. This indicates that a small amount of fast-moving material persists even at relatively large angles from the jet axis. The presence of this high-velocity material in the jet’s wings is likely related to the complex jet structure presented in Section \ref{sec:2shell}.

At larger angles ($\Theta\!\gtrsim\!20^\circ$), the outflow becomes dominated by sub-relativistic material ($u\lesssim 1$), and the velocity distribution becomes increasingly skewed towards low $u$. Nevertheless, small amounts of faster material are still present at these wide angles, though their energy content is comparatively small.

The above results suggest that these jets have a high degree of angular and velocity stratification, with a fast, collimated core surrounded by slower, less energetic wings. The transition between these regions is smooth rather than abrupt, challenging simplistic `top-hat' and `two component' jet models and highlighting the possibility of long-lasting energy injection in the external shock at all angles due to the broad velocity profiles, pointing to the need for more nuanced, structured outflows in afterglow modelling.

\begin{figure}
    \centering
    \includegraphics[width=\linewidth]{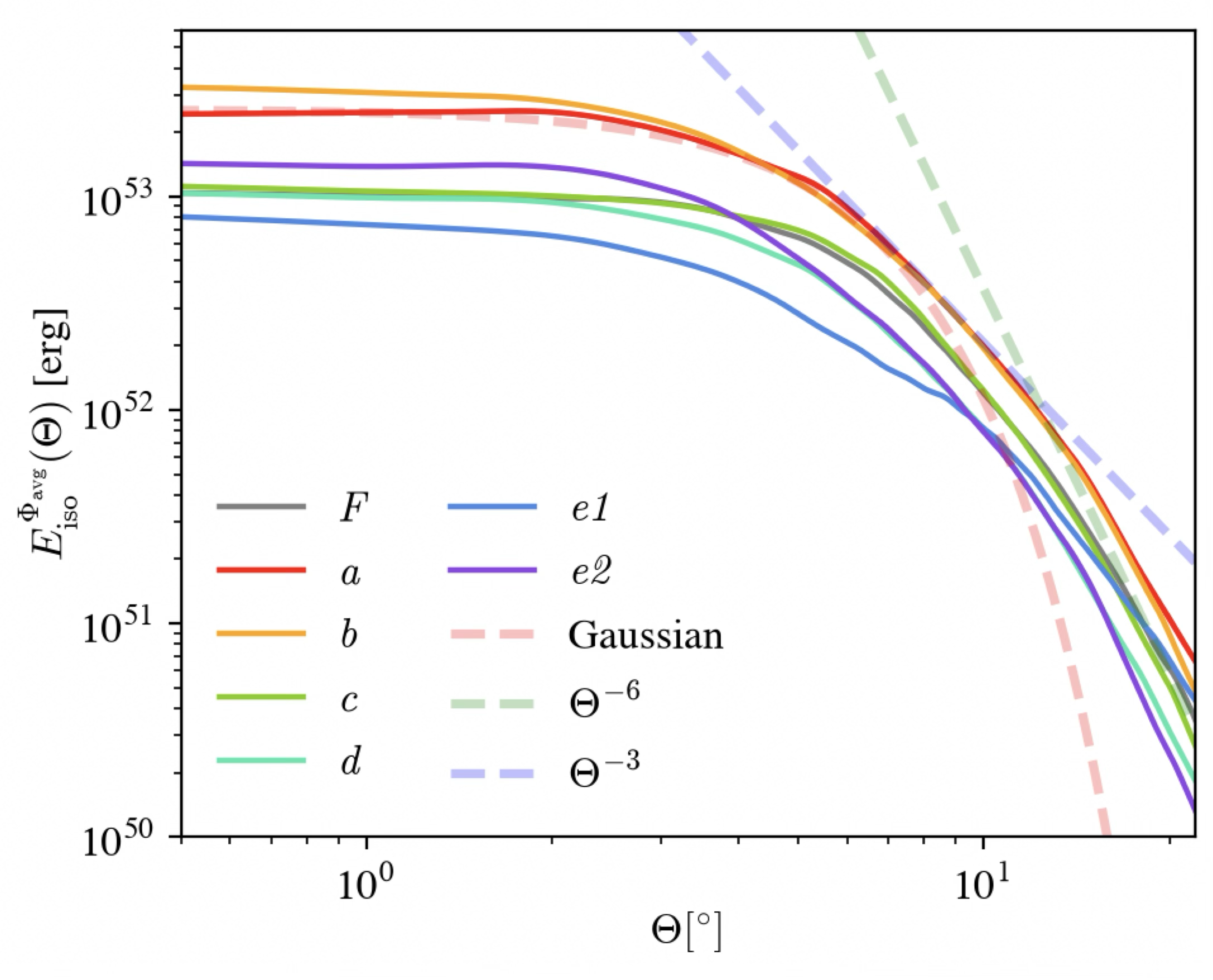}
    \caption{$E_\mathrm{iso}^{\Phi_\mathrm{avg}}(\Theta)$ profiles for all jet models at $t_\mathrm{end}$.}
    \label{fig:1d_eiso}
\end{figure}

To facilitate the comparison with analytical jet structures, in Fig. \ref{fig:1d_eiso} we show the azimuthally averaged $E_\mathrm{iso}^{\Phi_\mathrm{avg}}(\Theta)$ profiles at $t_\mathrm{end}$ for all the models. All jet simulations feature a flat energetic core for $\Theta \leq 2-3^\circ$. The less energetic wings are characterised by multiple segments with different slopes, affecting the rising part of the afterglow light curves for off-axis observers (see Section \ref{sec:af_model}).

Immediately outside the core, most models follow an approximate power-law decay
$E_{\rm iso}(\Theta)\propto \Theta^{-a}$ with $a \approx 3$; we show this slope in the figure purely as a visual reference for this segment. At larger angles, $\Theta \gtrsim 10^\circ$, the wings steepen: most models follow $a \approx 6$, while model \textit{e1} shows a shallower decay, $a \approx 4$, consistent with its broader, less collimated structure. A Gaussian profile, $E_{\rm iso}(\Theta) = E_{\rm iso}(0) \exp\left[ -\frac{\Theta^2}{2\theta_c^2}
\right],$ with $\theta_c = 4^\circ$, normalised to the peak of model \textit{a}, is also shown in the figure for comparison. We conclude that the jet structures produced by our simulations are complex and articulated, and do not generally follow a simple power-law or Gaussian functional form.

\section{Afterglow light curves}
\label{sec:afterglow}

\subsection{The model}
\label{sec:af_model}

To compute the broadband afterglow emission from our simulated jets, we employ a semi-analytical model based on  \citet{Salafia2019,Salafia2022b}, tailored for relativistic outflows with both angular and velocity structure. The model computes synchrotron emission from shock-accelerated electrons in a decelerating relativistic blast wave. The version of the model used in this work introduces two critical updates to better reflect the physical complexity revealed by our 3D simulations.
First, the code has been extended to incorporate the radial velocity profile of the jet extracted from the simulation data, as described in Appendix \ref{sec:blaswave_dynamics}. Instead of treating the outflow as a uniform shell with a single Lorentz factor at each angle, we consider the distribution of energy across fluid elements with different four-velocities, $u$. This allows us to model energy injection from slower shells catching up with the shock front, a process that is especially important during the rising phase of the afterglow light curve. 
Second, the assumption of axisymmetry is relaxed. Light curves are computed directly from the full 3D energy and velocity distributions, including azimuthal variations. This is essential for capturing the observational signatures of asymmetric outflows, which can differ substantially from their azimuthally averaged counterparts, particularly for off-axis observers.

Together, these upgrades allow us to more accurately predict the emission from realistic GRB jets and to explore the consequences of internal structure and asymmetry on the observed light curves.
We caution, though, that the model does not include the effects of lateral expansion of the shock, which are likely relevant at late enough times. A more detailed discussion on the expected extent of this limitation can be found in section \ref{sec:lc_velprof}. 
We further note that the light curves presented below are computed from jet structures obtained at a fixed numerical resolution, and that some of the resulting features depend on this choice, as discussed in Appendix~\ref{app:resolution}.

\begin{table*}[!t]
\caption{Result of MCMC fitting of each model to GW170817 afterglow data.}
\label{tab:MCMC}
\centering
\setlength{\tabcolsep}{4.3pt}
\renewcommand{\arraystretch}{1.3}
\begin{tabular}{ccccccccc}
\hline\hline
Parameter & Prior$^{(a)}$ & \multicolumn{7}{c}{Fit results$^{(b)}$} \\
 & & \textit{F} & \textit{a} & \textit{b} & \textit{c} & \textit{d} & \textit{e1} & \textit{e2} \\
\hline
$\log(n/\mathrm{cm^{-3}})$ & $\mathcal{U}(-5,-1)$ & $-1.99^{+0.3}_{-0.39}$ & $-1.53^{+0.40}_{-0.39}$ & $-1.69^{+0.48}_{-0.46}$ & $-2.05\pm0.39$ & $-2.41\pm0.35$ & $-1.69^{+0.29}_{-0.27}$ & $-2.46^{+0.37}_{-0.39}$ \\
$\log(\epsilon_\mathrm{B})$ & $\mathcal{U}(-8,-1)$ & $-3.49\pm0.12$ & $-4.25\pm0.12$ & $-4.24^{+0.13}_{-0.15}$ & $-3.53^{+0.12}_{-0.13}$ & $-3.16\pm0.12$ & $-3.05^{+0.10}_{-0.09}$ & $-3.33\pm0.12$ \\
$\theta_\mathrm{v}/\mathrm{deg}$ & Isotropic & $32.0^{+3.3}_{-3.4}$ & $32.9^{+3.4}_{-3.5}$ & $30.9^{+3.8}_{-3.7}$ & $31.3^{+3.3}_{-3.1}$ & $28.7^{+2.9}_{-2.5}$ & $38.4^{+2.8}_{-2.7}$ & $27.0^{+2.8}_{-2.7}$ \\
\hline
\end{tabular}
\tablefoot{
\tablefoottext{a}{`$\mathcal{U}(a,b)$' means uniform between $a$ and $b$; `Isotropic' means uniform in the cosine of the angle.}
\tablefoottext{b}{Median and 90\% credible interval of the marginalised posterior probability of each parameter for each of the models.}
}
\end{table*}

\subsection{Choice of reference parameters}
\label{sec:af_param}

In order to compute the afterglow light curves from the blastwaves caused by our simulated outflows, we need to specify a number of additional parameters: the number density $n$ of the external medium (assumed homogeneous); the `micro-physical' parameters that describe the fraction of the shock downstream internal energy in relativistic electrons ($\epsilon_\mathrm{e}$), the fraction in turbulent magnetic field ($\epsilon_\mathrm{B}$) and the slope of the power law energy distribution of the accelerated relativistic electrons ($p$); the redshift $z$ and luminosity distance $d_\mathrm{L}$ of the source; the polar and azimuthal viewing angles, $\theta_\mathrm{v}$ and $\phi_\mathrm{v}$. 

In order to compare the light curves with observations we focus on the afterglow associated to GW170817. Thus, we set the redshift to $z=0.0096$ and luminosity distance to $d_\mathrm{L}=40\,\mathrm{Mpc}$ \citep{Cantiello2018}. Moreover, we fix $\epsilon_\mathrm{e}=0.1$, based on theoretical expectations \citep{Sironi2013}, and we set $p=2.166$,  as constrained by multi-wavelength observations of the GW170817 synchrotron emission \citep{Fong2019}. This leaves us with four free parameters, namely $(n,\epsilon_\mathrm{B},\theta_\mathrm{v},\phi_\mathrm{v})$. 

We obtained reference values of $n$, $\epsilon_\mathrm{B}$ and $\theta_\mathrm{v}$ for each of our models by fitting the GW170817 data from \citet{Makhathini2021} adopting the axisymmetrised version of our jet structures at $t_\mathrm{end}$. The fit was performed through a Markov Chain Monte Carlo (MCMC) technique as implemented in the \texttt{emcee} python package \citep{Foreman-Mackey2013}. We assumed a log-uniform prior on $n$ in the range $(10^{-5},10^{-1})\,\mathrm{cm^{-3}}$, where the lower limit is informed by the smallest densities expected in galactic outskirts \citep{Fong2015}, while the upper limit is informed by observations of the GW170817 environment \citep{Hajela2019}; we set the prior on $\epsilon_B$ as log-uniform in the range $(10^{-8}, 10^{-1})$. The upper end is informed by theoretical expectations from particle-in-cell (PIC) simulations of relativistic shocks \citep{SironiSpitkovsky2011, Sironi2013}, while the lower end accounts for the possibility that the micro-turbulent magnetic fields generated at the shock decay as they are advected downstream, leading to effective values of $\epsilon_B$ well below the immediate post-shock level measured in PIC simulations. This picture, formalised by \cite{Gruzinov2001,MilosavljevicNakar2006} and \cite{Lemoine2013}, 
predicts a spatially decaying $\epsilon_B$ that reconciles the $\sim 10^{-3}$ values found at the shock in simulations, with the much lower values ($\sim 10^{-5}$--$10^{-7}$) inferred from broadband afterglow fits \citep{Santana2014,BarniolDuran2014,Ryan2024}. We set the prior on $\theta_\mathrm{v}$ as isotropic, that is, uniform in $\cos\theta_\mathrm{v}$ between 0 and 1.

The results of the fitting procedures are reported in Table~\ref{tab:MCMC}, and the 3\,GHz best-fit light curve\footnote{The best-fit models are obtained using the full set of available multi-band observations of GW170817. We find that our conclusions are consistent across comparisons with data in the optical, X-ray, and radio bands. Since the radio band provides the most extensive data coverage, for simplicity all light curves presented in this work are computed at a frequency of 3\,GHz.} of model \textit{F} is shown in Fig. \ref{fig:lc_sym_OFF}  (see Appendix~\ref{sec:curves} for the other models). Corner plots of the posterior probability distributions are also shown in Appendix~\ref{sec:curves} (Figs.~\ref{fig:corner_Fab} and \ref{fig:corner_cde1e2}).
We stress that the purpose of the MCMC fitting procedure is to obtain reference values for the afterglow parameters, anchoring the light curve computation to a well-observed event, rather than to provide a definitive physical model of GW170817; the recovered parameters should accordingly be interpreted as representative of the region compatible with each jet structure and not as unique estimates.
%
The MCMC returns a general preference for low values of $\epsilon_B$ ($\sim 10^{-3}$--$10^{-4}$), combined with relatively high values of $n$ (in some cases $\gtrsim 0.01$\,cm$^{-3}$). This combination reflects the well-known $n$--$\epsilon_B$ degeneracy in afterglow fitting (e.g., \citealt{granot2014}): a lower magnetic fraction in the shock downstream is compensated by a higher ambient density in reproducing the observed flux level. The best-fit values reported in Table~\ref{tab:MCMC} should accordingly be regarded as representative of the region of the $(n, \epsilon_B)$ plane compatible with the data given each jet structure; a broader discussion of this point is deferred to Section~\ref{sec:conclusion}. Our inferred values of $\epsilon_B$ are consistent with the theoretical framework discussed above, and the inferred densities remain within the prior range informed by observations of the GW170817 environment ($n < 10^{-1}$\,cm$^{-3}$, \citealt{Hajela2019}).
The best-fit viewing angles $\theta_v$ fall in the range between 27 and 33 degrees, with the exception of model \textit{e1} (giving about 38 degrees). These values lie outside of the range inferred from VLBI observations of the apparent superluminal motion of the afterglow centroid ($\theta_v \sim 19$--$25^\circ$; \citealt{Mooley2018a,Ghirlanda2019,Mooley2022}). 
We note that the centroid motion is not included as a constraint in our MCMC, since incorporating it would steer the fits toward reproducing the specifics of GW170817, which is not the aim of the present analysis. 

\subsection{Impact of velocity profile}
\label{sec:lc_velprof}

\begin{figure}
    \centering
    \includegraphics[width=\columnwidth]{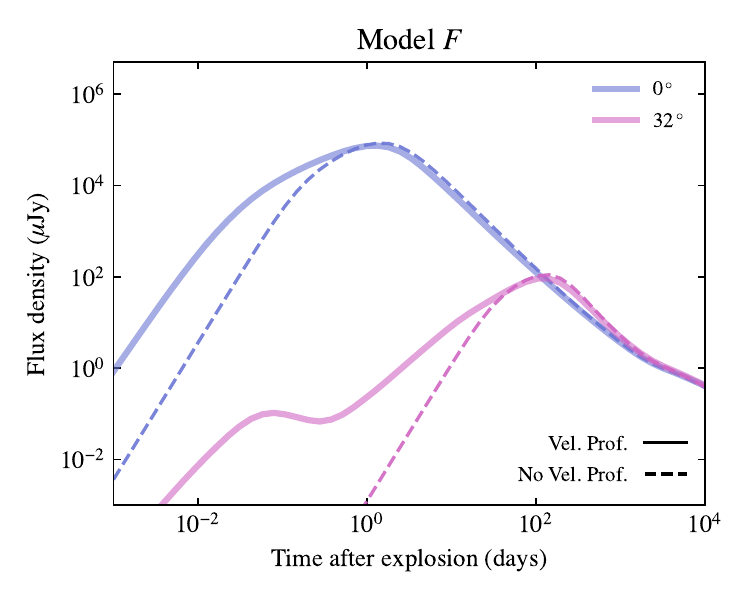}
    \caption{Afterglow light curve at 3 GHz for the fiducial case at $t_\mathrm{end}$, with or without the inclusion of the velocity profile in the afterglow model, for both an on-axis and an off-axis observer.} 
    \label{fig:lc_velprof}
\end{figure}

In Fig. \ref{fig:lc_velprof} we compare afterglow light curves for the fiducial model \textit{F} computed with and without the inclusion of the velocity profile, adopting the values of $n$, $\epsilon_\mathrm{B}$ and $\theta_\mathrm{v}$ best-fitting GW170817 afterglow data (see Table~\ref{tab:MCMC}). In addition, we consider the on-axis case ($\theta_\mathrm{v}\!=\!0$). Here and in the following, we show in particular the light curves obtained at 3\,GHz.
We recall that, in the case without the velocity profile, all material at a given angle is assigned a single Lorentz factor, corresponding to an `average' saturation value computed from the total energy of the outflow. At this stage, we still refer to the azimuhtally averaged jet profiles. The assumption of axisymmetry will be relaxed in Section (\ref{sec:lc_asymmetry}).

The most immediate effect of including the velocity profile is a broadening of the light curve, especially in its early rising phase. The sharp rise seen in the simplified case becomes more gradual, as slower inner shells catch up with the forward shock, injecting energy over time. This is a natural consequence of the stratified structure seen in the simulations, where the energy is distributed over a relatively wide range of four-velocities even near the jet axis (see Fig.~\ref{fig:vel_theta}).

Additionally, the quasi-spherical outer shell accompanying the expanding jet (Section~\ref{sec:2shell}) gives rise to a secondary, dimmer peak at early times, visible at off-axis viewing angles. Due to its intrinsically low luminosity and early occurrence, this feature is likely difficult to observe. A dedicated discussion is presented in Appendix \ref{sec:shell_detectability}.

As mentioned in Section~\ref{sec:af_model}, the velocity profile and a standard treatment of lateral expansion cannot currently be combined self-consistently in our model, but we can assess their impact separately for the fiducial model, relative to the common baseline of not including either of the two effects. Including the velocity profile, our default choice, broadens the rising phase and, for off-axis observers, introduces the early, dim precursor associated with the outer shell, while leaving the on-axis peak nearly unchanged. Lateral expansion alone has a negligible effect on the rising part; its main influence is a significant steepening of the decay beyond the peak.
Off-axis, it also mildly regresses the peak and slightly affects its flux. We therefore expect a combination of the two ingredients to further steepen the decay with respect to the velocity-profile-only case shown here. We caution, however, that these effects are likely mitigated in presence of a broad velocity profile: this happens because lateral gradients across the shock are smoothed out at the local speed of sound (as measured in the shock downstream), which diminishes as the shock decelerates. As a result, the anisotropy caused by late-time energy injection (which can be substantial in presence of a broad velocity profile) is likely preserved to a larger extent with respect to the case where all the energy is injected in the shock at early times \citep[similar arguments are laid out by][]{Sadeh2025}. We leave a joint implementation and a quantification of these effects for future work.

\subsection{Light curves at different evolution times}
\label{lc_time}

To assess the importance of extending the jet simulations beyond the initial evolutionary phase, we compare afterglow light curves computed from outflow data at different post-launch times. Specifically, we contrast results obtained at $t-t_\mathrm{j}=3$\,s, the endpoint of the jet simulation presented in \citetalias{Pavan2025}, with those at the end of our extended 3D simulation (i.e.~at $t_\mathrm{end}=9.5$\,s), 
The goal of this comparison is to understand how the continued evolution of the outflow affects observable quantities. During the early post-breakout phase, the jet undergoes significant reorganisation: residual acceleration, cocoon expansion, and mixing can all reshape the angular and velocity structure of the outflow. These effects are not captured in models that truncate shortly after breakout.

\begin{figure}
    \centering
    \includegraphics[width=\linewidth]{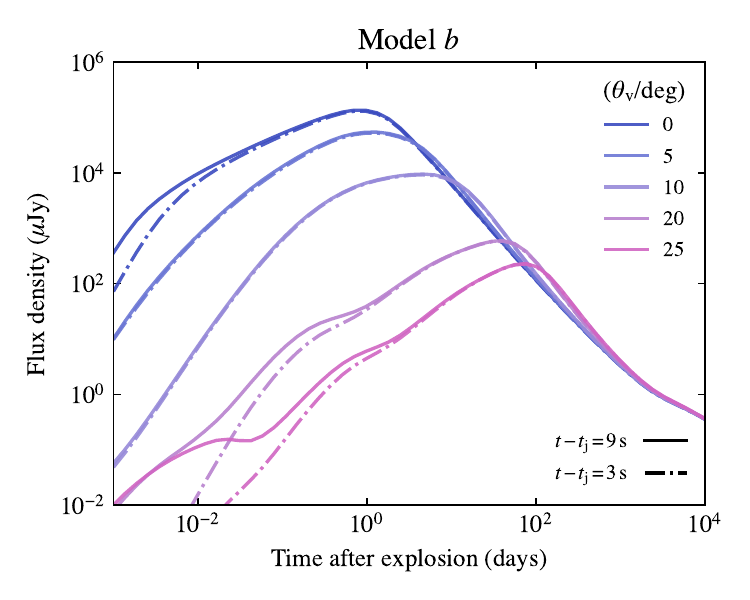}
   \caption{Afterglow light curve at 3 GHz for model \textit{b} at the beginning and at the end of the extended evolution, considering a set of viewing angles $\theta_\mathrm{v}$}.
    \label{fig:lc_times}
\end{figure}

In Fig.~\ref{fig:lc_times}, we show such a comparison for model $b$, considering the azimuthally averaged jet profiles and different viewing angles $\theta_\mathrm{v}$.
Our results show that the later-time simulations predict noticeable differences in the early rising phase of the light curve. In fact, while the late-time behaviour (decay slope) is identical across the two snapshots, the early light curve is sensitive to the detailed structure at different evolutionary times.

This confirms the value of extending jet simulations to at least $\sim\!10$\,s, especially for interpreting the rising phase of the afterglow, which is crucial for constraining jet geometry, energy, and viewing angle.
\begin{figure}[b!]
    \centering
    \includegraphics[width=0.95\linewidth]{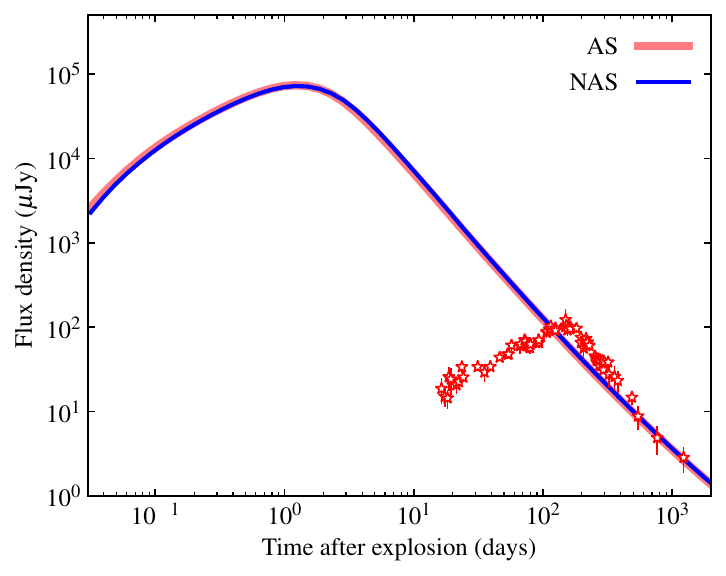}
    \caption{On-axis afterglow light curves at 3 GHz computed using the axisymmetrised (AS -- thick red line) and not axisymmetrised (NAS -- blue line) jet profiles from model $F$. Red stars with vertical error bars correspond to the GW170817 afterglow data.}
    \label{fig:lc_symm_onax}
\end{figure}
\begin{figure}
    \centering
    \includegraphics[width=1\linewidth]{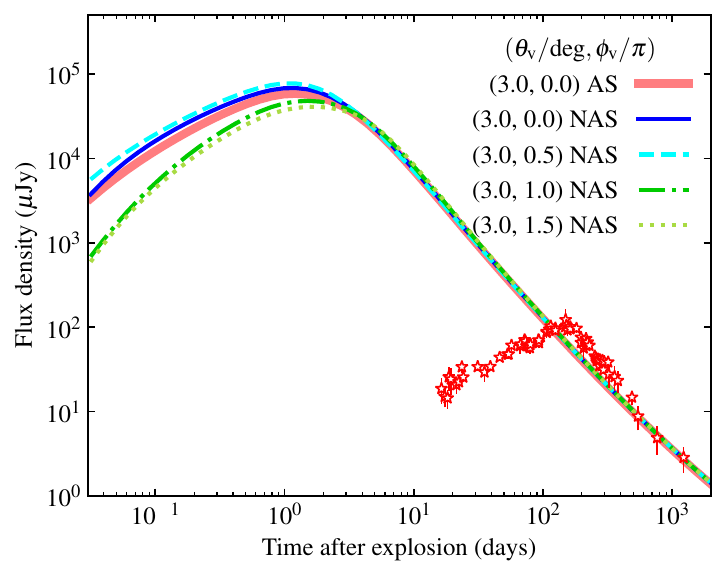}
    \caption{Analogous to Fig.~\ref{fig:lc_symm_onax} for slightly off-axis observation angle ($\theta_\mathrm{v} = 3^\circ$). In the non-axisymmetrised (NAS) case, we also vary the azimuthal viewing angle $\phi_\mathrm{v}$.}
    \label{fig:lc_symm_off}
\end{figure}
\begin{figure}
    \centering
    \includegraphics[width=1\linewidth]{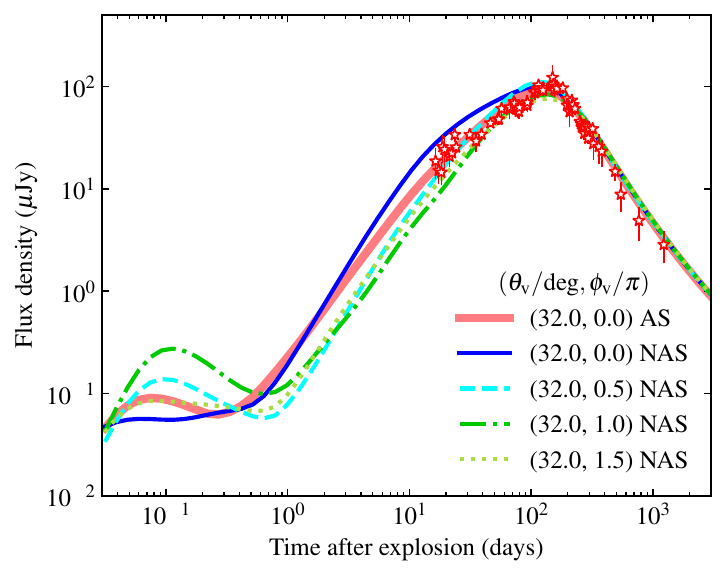}
    \caption{Analogous to Fig.~\ref{fig:lc_symm_off} for $\theta_\mathrm{v} = 32^\circ$.} 
    \label{fig:lc_sym_OFF}
\end{figure}

\subsection{Impact of azimuthal asymmetries}
\label{sec:lc_asymmetry}

Here, we investigate the impact of azimuthal asymmetries in the jet structure on the resulting afterglow light curves. The extent and origin of such deviations from axisymmetry is discussed in Section~\ref{sec:angular}.

In Fig. \ref{fig:lc_symm_onax}, we show light curves for the fiducial model and an on-axis observer ($\theta_\mathrm{v}\!=\!0$), comparing results obtained from the axisymmetrised (`AS') and non-axisymmetrised (`NAS') jet structure. 
We observe negligible difference between the two curves. 

In Fig.~\ref{fig:lc_symm_off}, we compare the axisymmetrised and non-axisymmetrised cases for a slightly off-axis observer ($\theta_\mathrm{v} = 3^\circ$) and at four different azimuthal viewing angles $\phi_\mathrm{v}=(0, 1, 2, 3)\times \pi/2$, evenly spaced around the jet axis so as to sample representative orientations with respect to the non-axisymmetric jet structure.
This allows us to highlight the effects of structural asymmetries. We find that varying the azimuthal viewing angle can produce order-of-magnitude differences in the rising part and the peak of the light curve, which then decrease after the peak and become negligible (this is even more evident for model \textit{a}, as shown in Fig.~\ref{fig:symm_all}).

In Fig.~\ref{fig:lc_sym_OFF} we perform a similar analysis but for a strongly off-axis observer ($\theta_\mathrm{v}\!=\!32^\circ$, i.e.~the best-fitting value with respect to GW170817 data; see Table~\ref{tab:MCMC}). 
Also in this case, varying the azimuthal angle produces large differences up to the peak time.
We find that the early-time peak due to the multiple-shock structure, for off-axis observers, is strongly dependent on the azimuthal viewing angle. 
This indicates that the energetic shells responsible for the early-time emission are not uniformly distributed along the azimuthal direction. 

Finally, we performed MCMC best-fit analysis for model \textit{F} (see Section \ref{sec:af_param}) by employing the non-axisymmetrised jet structure, for three different azimuthal viewing angles, i.e. $\phi_v = 0, \pi/2, \pi$ (corner plots are shown in Appendix \ref{sec:curves}, Fig.~\ref{fig:corner_Fview}). We find substantial differences in the parameter estimations with the different azimuthal viewing angles and with respect to the axisymmetrised case. Focusing on the estimation of the polar viewing angle, we find 
$\theta_v \simeq 41.6^\circ \pm 3^\circ, 26.3^\circ \pm 3^\circ, 23.9^\circ \pm 3^\circ$ for the three choices of $\phi_v$, respectively, spanning a range of about $18^\circ$, larger than the statistical uncertainty of any individual fit. In particular the estimate of $\theta_v$ is not compatible with the one obtained from the axisymmetrised structure (within 90\% credible intervals). Similar shifts appear in the estimates of $\epsilon_B$ and $n$: the non-axisymmetrised fits yield systematically higher $\log \epsilon_B$ (ranging from $-2.94 \pm 0.12$ to $-2.41 \pm 0.17$, compared to $-3.49 \pm 0.12$ in the axisymmetrised case), while $\log n$ takes the values $-1.58 \pm 0.33, -3.20 \pm 0.45, -3.36 \pm 0.50$ for the three $\phi_v$, respectively, bracketing the axisymmetrised value of $-1.99 \pm 0.40$ from both sides.
This confirms that the azimuthal orientation of the observer with respect to the jet structure, unknown and effectively random for any given event, is a genuine and commonly overlooked source of systematic uncertainty on the inferred parameters. The four $\phi_v$ orientations sampled here should be regarded as an estimate of the possible variations rather than a statistical characterisation; nonetheless, the resulting order-of-magnitude spread in the pre-peak flux and the $\sim 18^\circ$ spread in the inferred $\theta_v$ already indicate that the systematic uncertainty associated with this effect is substantial.

The full set of light curves for all jet models is presented in Appendix~\ref{sec:curves}.

\subsection{Impact of jet injection parameters on afterglow light curves}
\label{sec:fits}

\begin{figure}
    \centering
    \includegraphics[width=0.95\columnwidth]{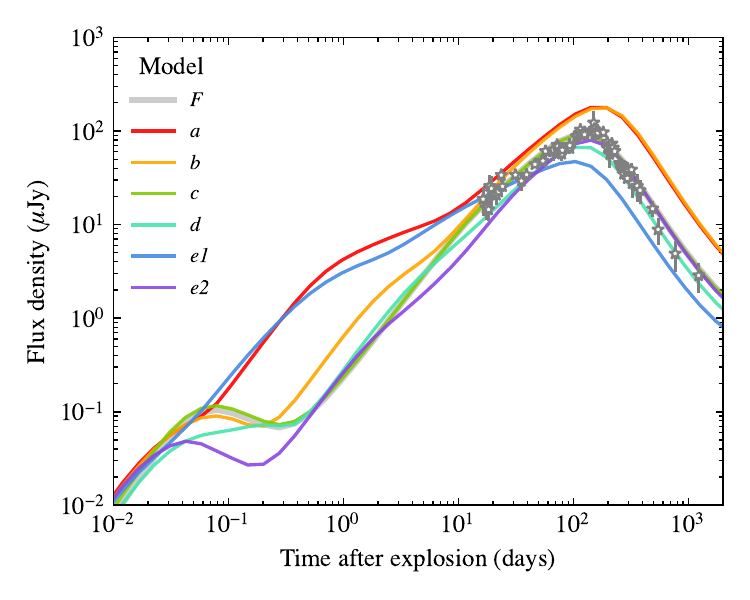}\\
    \includegraphics[width=0.95\columnwidth]{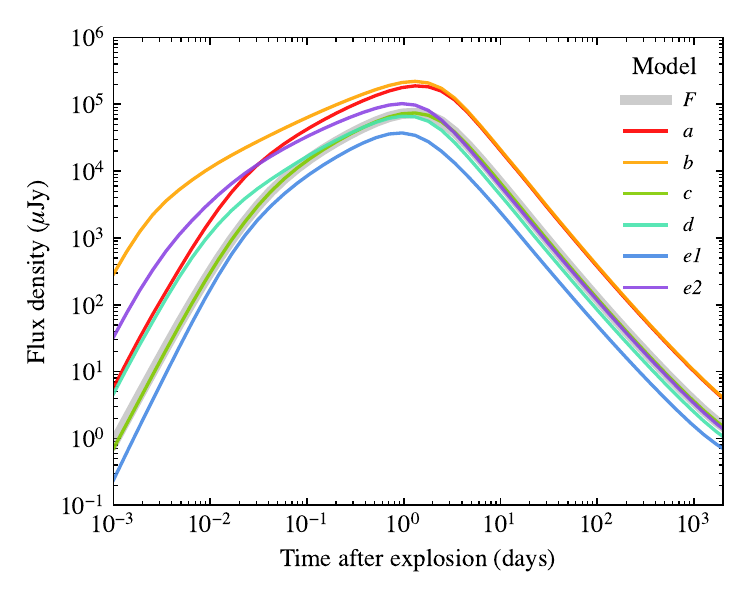}\\
    \caption{Top: Comparison of afterglow light curves at 3 GHz obtained from the axisymmetrised jet structures of  the different models, adopting a fixed set of afterglow parameters (namely, the best-fit values obtained for the fiducial model $F$, see Table~\ref{tab:MCMC}). Grey stars with vertical error bars correspond to GW170817 data. 
    Bottom: Analogous comparison for the on-axis observer ($\theta_\mathrm{v}=0$).}
    \label{fig:lc_jets}
\end{figure}

In this Section, we assess how the representative set of jet injection variations introduced in Section \ref{sec:jet_params} shape the resulting afterglow light curves. Figure \ref{fig:lc_jets} compares the light curves obtained from the axisymmetrised versions of jet structures from the different models, computed with a fixed set of afterglow parameters equal to the best-fit values of the fiducial case. We show the results for both on- and off-axis observers.

The rising phase of the off-axis light curves (shown in the top panel of the Figure) is primarily controlled by the jet’s angular structure (Figure \ref{fig:1d_eiso}) and the presence of a velocity profile (Section~\ref{sec:lc_velprof}). Steeper energy distributions outside the jet core lead to a correspondingly steeper rise, as progressively more energy enters the observer’s line of sight when the jet decelerates and the relativistic beaming cone widens. Signatures of changes in slope in the energy distribution and the presence of different components in the velocity profile are directly imprinted on the light curves as breaks or slope changes during the rise.

The time scale $t_\mathrm{peak}$ of peak emission scales with the jet energy through the Sedov length $R_\mathrm{S}$ (i.e.~the fundamental length scale of the problem, e.g.\ \citealt{Blandford1976}), that is $t_\mathrm{peak}\!\propto\! R_\mathrm{S}/c\propto [E_\mathrm{iso}(\Theta=0)/n]^{1/3}$. More energetic and collimated jets (models \textit{a}, \textit{b}, and \textit{e2}) peak indeed at later times. As seen from both the on- and off-axis curves (bottom and top panel of Figure \ref{fig:lc_jets}, respectively), their peaks are also brighter, since a larger $R_\mathrm{S}$ 
corresponds to a greater number of emitting electrons.

Models \textit{c} and \textit{d} differ very little from the fiducial case, as expected due to the minor difference in luminosity and/or magnetisation.

Comparing models \textit{e1} and \textit{e2}, which differ in jet launching time, we observe that \textit{e2} produces a more typical afterglow evolution with a brighter and delayed peak, while \textit{e1} presents an extended plateau prior to peak time, produced by its peculiar energetic structure. 
Model \textit{e1} has indeed a less energetic, less collimated, and more off-centered jet core; we note that this off-axis plateau is however a feature found to depend on numerical resolution (see Appendix~\ref{app:resolution}).

At late times, the decaying phase of the light 
curve is no longer sensitive to the jet structure, and the models converge to similar slopes.

\section{Summary and outlook}
\label{sec:conclusion}

In this work, we presented a set of 3D relativistic MHD simulations of jets emerging from a realistic binary neutron star merger environment, extending the work presented in \citepalias{Pavan2025} to reach saturation in terms of jet structure and acceleration. 
Then, we used the output of our simulations to compute the afterglow emission of the different jets with an improved semi-analytic code, assessing the impact of the differences arising in the jets' structure with time and angle. Finally, we performed MCMC fits to the data of the GW170817 event, obtaining the best-fit values for a subset of the afterglow model parameters.

From the dynamical evolution of the various jet models, given the specific choice for the input parameters we find that more energetic jets experience a more successful drilling through the merger environment, losing less energy with respect to the fiducial case. As a consequence, especially for 
a longer decay time scale for injection, the resulting jet structure is only slightly displaced from the injection axis and exhibits a more energetic core.
The launching time has also major impact on the jet evolution, with jets launched earlier resulting in a more axisymmetric, collimated, and energetic core.
Finally, increasing the magnetisation helps maintaining a more stable jet-environment interface, preventing the development of instabilities. Nevertheless, given the choice made for the input parameters, we do not find major variations in the evolution and final structure of the jet models with different magnetisation.
Despite the distinction between the various models in the pre-breakout phase, which primarily influence the resulting jet energetics and structure, they all behave similarly in the late time evolution. As previously shown in \citetalias{Dreas2025}, we find that the 3D evolution within a realistic post-merger environment leads to asymmetries (i.e.~deviations from axisymmetry) in the jet structure, especially for the weaker jets that are more affected by this interaction.
Notably, we find that at the time of the jet-cocoon breakout, all models develop a complex multiple-shock structure, with an outer shell of non negligible energy preceding the jet ejecta, which is responsible for producing an early time dimmer peak in the afterglow light curves.


Turning from the dynamical simulations to their afterglow implications, and using GW170817 as a reference case to anchor the afterglow model, we performed MCMC fits adopting the axisymmetrised jet structures of each model. 

All the jet structures considered can be brought into reasonable agreement with the observational data within the explored parameter space. 
The best-fit values favour polar viewing angles between 27 and 33 degrees (except for the model with the lowest injected energy and magnetisation, giving about 38 degrees), low $\epsilon_B$ and relatively high $n$.
We note that our results are robust across comparisons with optical, X-ray, and radio observations of GW170817, and therefore we focus on the radio band, at 3 GHz, due to its superior data coverage. Nevertheless, multi-band comparisons remain essential, particularly for future events where improved observational coverage across different wavelengths may provide stronger constraints.

To assess the impact of the different jet injection prescriptions in shaping the emission, we compare the light curves from the different configurations for a fixed set of afterglow parameters, corresponding to the best-fit values obtained for the axisymmetrised jet structure of the fiducial model. This comparison is made for both on-axis and off-axis observers. 
As expected, more energetic jets produce a brighter emission, peaking at later times. In particular, we find a difference of 1.5 orders of magnitude in the on-axis peak emission between the least and the most energetic jet models. 
We conclude that variations in the initial conditions of jet launching produce differences in the resulting afterglow light curves, both in peak flux and in the temporal evolution, that reflect the distinct jet structures (Section~\ref{sec:af_param}). 

The impact on the afterglow emission of the azimuthal asymmetries is investigated by relaxing the assumption of axisymmetry in the afterglow model, and comparing the resulting emission by varying both $\theta_\mathrm{v}$ and $\phi_\mathrm{v}$.
While the on-axis emission remains nearly identical, for off-axis observers ($\theta_\mathrm{v}$ above a few degrees) we find up to an order-of-magnitude differences in the flux density for different choices of $\phi_\mathrm{v}$. The effect of the specific radial and angular structure of the different jets (including also the outer shell mentioned above) is properly taken into account only if the jet's velocity profile is included in the afterglow dynamics. 

The relevance of extending the simulations up to 9.5 seconds, when acceleration is mostly over and the angular structure is no longer changing, is mostly evident in the early-time part of the light curves, especially for large viewing angles.

The quantitative results presented above should be interpreted within the scope of applicability of our analysis. 
First, all simulations are performed at a fixed numerical resolution and are not yet in a convergence regime. 
As shown in our resolution study (Appendix~\ref{app:resolution}), performed for models \textit{e1} and \textit{e2}, indicates that qualitative trends and the ordering among models are preserved across resolutions, while off-axis some features, most notably the extended pre-peak plateau of model \textit{e1}, are more sensitive to this choice.
Accordingly, the quantitative predictions of this work should be read as indicative of the behaviour of each configuration, and differences between closely spaced models comparable to the resolution-induced variations should not be regarded as robustly resolved.
Second, the values of $n$ and $\epsilon_B$ returned by the MCMC fits (Table~\ref{tab:MCMC}) are subject to the well-known $n$–$\epsilon_B$ degeneracy in afterglow modelling (Sect.~\ref{sec:af_param}): other combinations of $n$ and $\epsilon_B$ lying along the degeneracy locus can reproduce the data comparably well, and the reported best-fit values should therefore be regarded as representative of the compatible region, given that additional independent constraints would be required to break the degeneracy. Third, the treatment of azimuthal asymmetries is based on four representative viewing angles $\phi_v = (0, 1, 2, 3)\times\pi/2$, which are sufficient to expose an order-of-magnitude spread in flux and a $\sim 18^\circ$ spread in the inferred $\theta_v$ (Sect.~\ref{sec:lc_asymmetry}). While this does not represent a statistical distribution over $\phi_v$, whose characterisation would require a computationally expensive number of realisations at the present resolution, it shows that structural asymmetries further increase the complexity of the problem by introducing additional effects in shaping the light curves. 


A direction for future development is to extend this study to higher spatial resolution, in order to better resolve the regions involved in the jet–environment interaction, through the implementation of adaptive mesh refinement (AMR) techniques, which make it possible to selectively enhance the resolution where it is most needed. This strategy would make the investigation of the multiscale dynamical evolution computationally tractable, while enabling the derivation of more robust and quantitatively reliable results. On this basis, extending the study to a broader range of injection parameters would allow a more systematic exploration of the regimes relevant for merger-driven events.
Our analysis can also benefit from improvements on the afterglow modelling side, such as a self-consistent combination of the lateral expansion with the radial velocity stratification within the afterglow model, and the inclusion of the radio centroid displacement among the observational constraints. The latter, which we do not employ in the present fits, provides complementary information to the light curve, and a dedicated modelling of the centroid motion may further restrict the set of jet structures compatible with the data. 
A possible extension of the angular analysis would be to study the effect of the structural asymmetries in terms of polarisation (see, e.g., \citealt{Lamb2022}). Moreover, considering the non-axisymmetrised jet structures, the azimuthal viewing angle can be taken into account as an additional parameter of investigation,
enabling a statistical characterisation of its impact on the inferred afterglow parameters that goes beyond the systematic estimate provided in this work.

\begin{acknowledgements}

We thank the anonymous referee for their careful reading and constructive comments, which helped to improve the quality of the study presented.
This work was supported by the European Union under NextGenerationEU, via the PRIN 2022 Projects “EMERGE: Neutron star mergers and the origin of short gamma-ray bursts", Prot. n. 2022KX2Z3B (CUP C53D23001150006), and “PEACE: Powerful Emission and Acceleration in the most powerful Cosmic Explosion”, Prot. n. 202298J7KT (CUP G53D23000880006). The views and opinions expressed are solely those of the authors and do not necessarily reflect those of the European Union, nor can the European Union be held responsible for them. 
We further acknowledge support from the INAF Theory Grant 2023 ``AfterJet: Modelling short gamma-ray burst jets from binary neutron star mergers and their afterglow emission'', Ob.Fu. 1.05.23.06.02 (CUP C93C23006800005).
AP and RC acknowledge additional support by the Italian Ministry of Foreign Affairs and International Cooperation (MAECI), grant number US23GR08. OS acknowledges financial support from INAF through grant 1.05.23.04.04.
Simulations were performed on the Discoverer HPC cluster at Sofia Tech Park (Bulgaria). We acknowledge EuroHPC Joint Undertaking for awarding us access to this cluster via the Regular Access allocation EHPC-REG-2023R03-160.
\end{acknowledgements}

\bibpunct{(}{)}{;}{a}{}{,} 
\bibliographystyle{aa} 
\bibliography{refs} 

\appendix

\section{Velocity profile}
\label{sec:velprof}

The velocity profiles extracted from our simulations provide crucial input for modelling the jet’s afterglow. 
Specifically, they allow us to track how much energy is carried by fluid elements at each angle and four-velocity. In the simplest jetted GRB models, all of the energy is uniformly distributed within a narrow angular region around the axis and is carried by ultra-relativistic material with a single, definite Lorentz factor. A more complex distribution of energy in the angle and in the Lorentz factor space leads to a more rich phenomenology, especially when the viewing angle is allowed to vary. 

\begin{figure}
    \centering
    \includegraphics[width=1\linewidth]{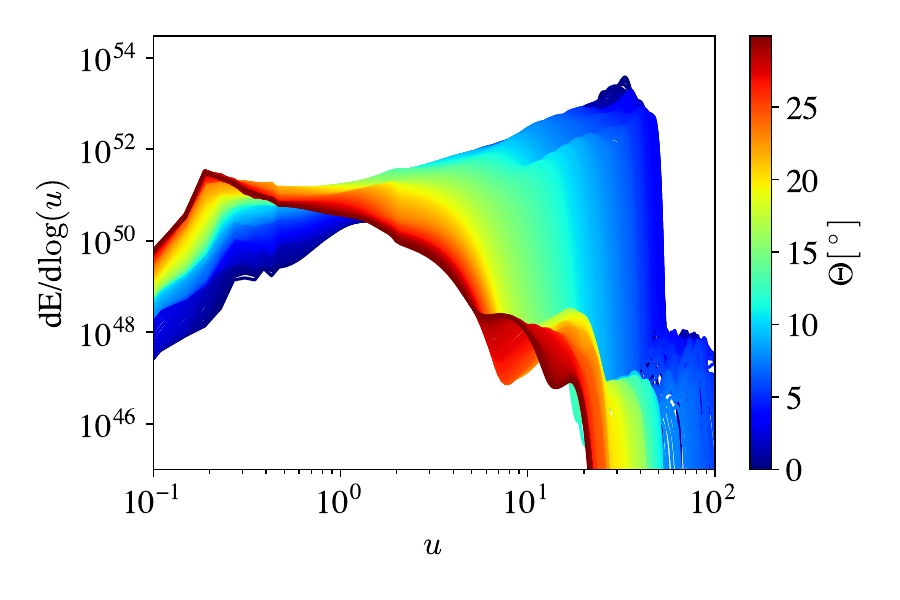}
    \caption{Velocity profile of the fiducial model for different polar angles.}
    \label{fig:velprof}
\end{figure}

Figure \ref{fig:velprof}, represents velocity profiles at different $\Theta$ angles (identified by different colours) for the fiducial model at $t_\mathrm{end}$, obtained from the azimuthally averaged (total) energy distribution. 
As already pointed out in Section~\ref{sec:angular}, we note that even at small angles from the injection axis the profiles are wide, indicating that a considerable amount of energy is carried by slower elements.

Moreover, we find that some high-velocity material exists even at large angles ($\Theta > 15^\circ$), albeit in small quantities. This component originates from the jet’s multiple-shell structure (see Section~\ref{sec:2shell}) and can produce observable signatures, particularly for off-axis observers (see Section~\ref{fig:lc_velprof}).

\section{Blastwave dynamics}\label{sec:blaswave_dynamics}

We compute the dynamics of the blastwave independently at each angular coordinate $(\Theta,\Phi)$ (therefore neglecting lateral energy transport) based on an approximate conservation of energy per unit solid angle. Let us assume that the isotropic-equivalent energy of the outflow $E_\mathrm{iso}$ is entirely in the form of kinetic energy plus rest mass energy. Let us then define the isotropic-equivalent energy in ejecta faster than $u$,
\begin{equation}
    E_\mathrm{iso}(>u) = \int_u^\infty \frac{\mathrm{d}E_\mathrm{iso}}{\mathrm{d}u}(u')\,\mathrm{d}u',
\end{equation}
the isotropic-equivalent rest mass in ejecta faster than $u$,
\begin{equation}
    m_\mathrm{ej,iso}(>u) = \int_u^\infty \frac{1}{c^2\sqrt{1+{u'}^2}}\frac{\mathrm{d}E_\mathrm{iso}}{\mathrm{d}u}(u')\,\mathrm{d}u'
\end{equation}
(where $c$ is the speed of light), and the isotropic-equivalent external medium mass swept by the blast wave when the four-velocity in its downstream is $u$, 
\begin{equation}
    m_\mathrm{iso}(u) = 4 \pi r^3(u) n m_\mathrm{p}/3,
    \label{eq:miso}
\end{equation}
where $n$ is the number density of the external medium (assumed to consist of ionised hydrogen), $m_\mathrm{p}$ is the proton mass, and $r(u)$ is the distance travelled by the shock up to the time when its downstream four-velocity is $u$. Our energy conservation equation reads \citep[cfr.][]{Panaitescu2000}
\begin{equation}
\begin{split}
& E_\mathrm{iso}(>u) + m_\mathrm{iso}(u) c^2 = \\
& \Gamma \left[m_\mathrm{ej,iso}(>u)+m_\mathrm{iso}(u)\right] c^2 + \Gamma_\mathrm{eff}\left[(\Gamma-1)m_\mathrm{iso}(u)\right]c^2
\end{split},
\label{eq:en_cons_afterglow}
\end{equation}
where $\Gamma=\sqrt{1+u^2}$,
\begin{equation}
    \Gamma_\mathrm{eff} = \frac{\hat\gamma_{\mathrm{ad}}\Gamma^2-\hat\gamma_{\mathrm{ad}}+1}{\Gamma}
\end{equation}
is the `effective Lorentz factor' that transforms the total internal energy from the fluid frame to the laboratory frame \citep{Nava2013}, and $\hat\gamma_\mathrm{ad}$ is the adiabatic index in the blastwave downstream, which we estimate following \citet{Pe'er2012}.
The assumptions in this formulation are: (i) the ejecta faster than the blastwave downstream velocity $u$ catch up immediately with it, injecting all their energy; (ii) the internal energy is dominated by that of freshly shocked external medium material (i.e.\ shocked over one adiabatic time scale, which corresponds roughly to one doubling in radius), which itself dominates in terms of mass over the material shocked at earlier time. Equation \ref{eq:en_cons_afterglow} can be solved analytically for $m_\mathrm{iso}(u)$, leading to
\begin{equation}
    m_\mathrm{iso}(u) = \frac{E_\mathrm{iso}(>u)/c^2-\Gamma(u) m_\mathrm{ej,iso}(>u)}{\left[\Gamma(u)-1\right]\left[\Gamma_\mathrm{eff}(u)+1\right]}.
\end{equation}
The above equation establishes a one-to-one relation between $m_\mathrm{iso}(u)$ and $u$, which can be turned into a relation between $r$ and $u$ using Eq.\ \ref{eq:miso}, namely $r(u)=[3 m_\mathrm{iso}(u) /4\pi n m_\mathrm{p}]^{1/3}$. The relation is monotonic and can be inverted numerically. This completely specifies the dynamics of the blastwave at each angular coordinate $(\Theta,\Phi)$.

\section{Light curves and fit results}
\label{sec:curves}
\begin{figure*}
    \centering
    \includegraphics[width=\textwidth]{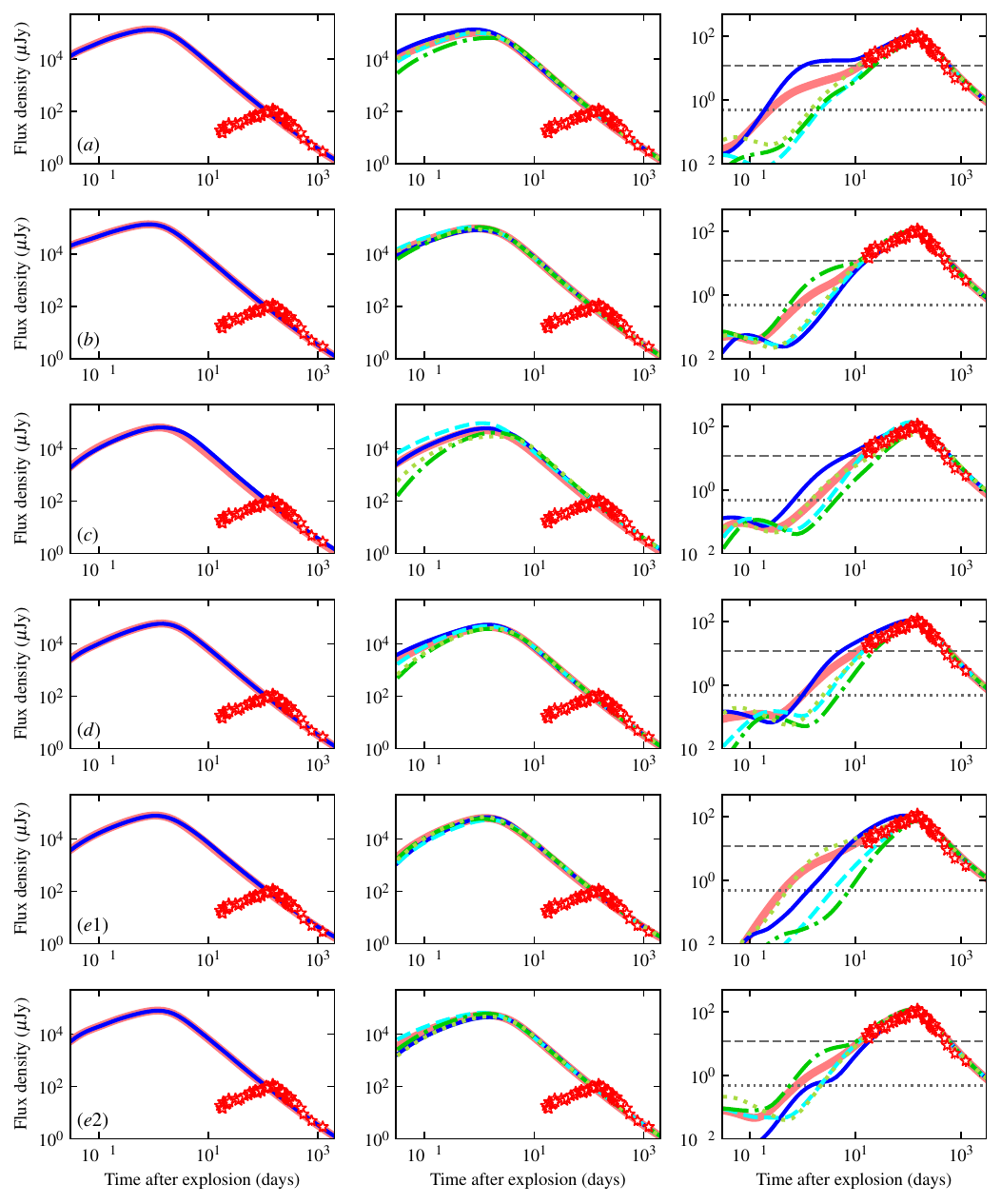}
    \caption{On-axis and off-axis light curves at 3 GHz comparing the axisymmetrised and non-axisymmetrised jet structures for all models, except for the fiducial one already presented in Section~\ref{sec:lc_asymmetry}.
    Each row refers to a different model (indicated in the lower-left corner of the left panel), with the left, middle and right panels being the equivalent of Figs.\ \ref{fig:lc_symm_onax}, \ref{fig:lc_symm_off} and \ref{fig:lc_sym_OFF}, respectively (also reflecting the same colours and line styles). 
    For each model, the polar viewing angle $\theta_\mathrm{v}$ in the rightmost panel is set to the best fit of that model, as reported in Table \ref{tab:MCMC}. Horizontal lines
mark the $3\sigma$ sensitivities of the VLA (dashed) and ngVLA (dotted) at
3 GHz for a fiducial 1 hr on-source integration.}
    \label{fig:symm_all}
\end{figure*}

In Figure \ref{fig:symm_all} we present the on-axis and off-axis light curves for all models, showing the effect of the jet's structural asymmetries, as discussed in Section~\ref{sec:lc_asymmetry}.

\begin{figure*}
    \centering
    \includegraphics[width=0.5\textwidth]{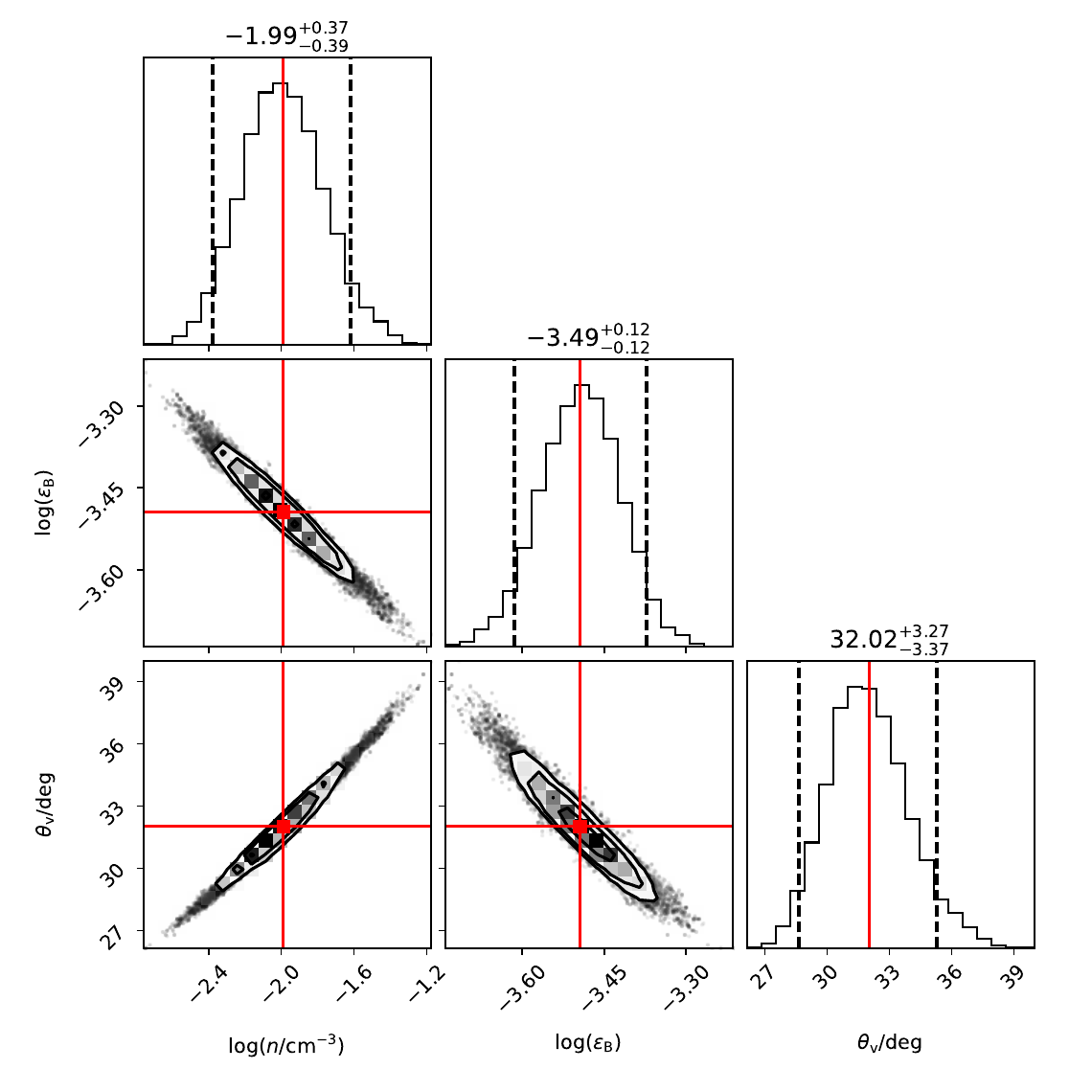}\\
    \includegraphics[width=0.5\textwidth]{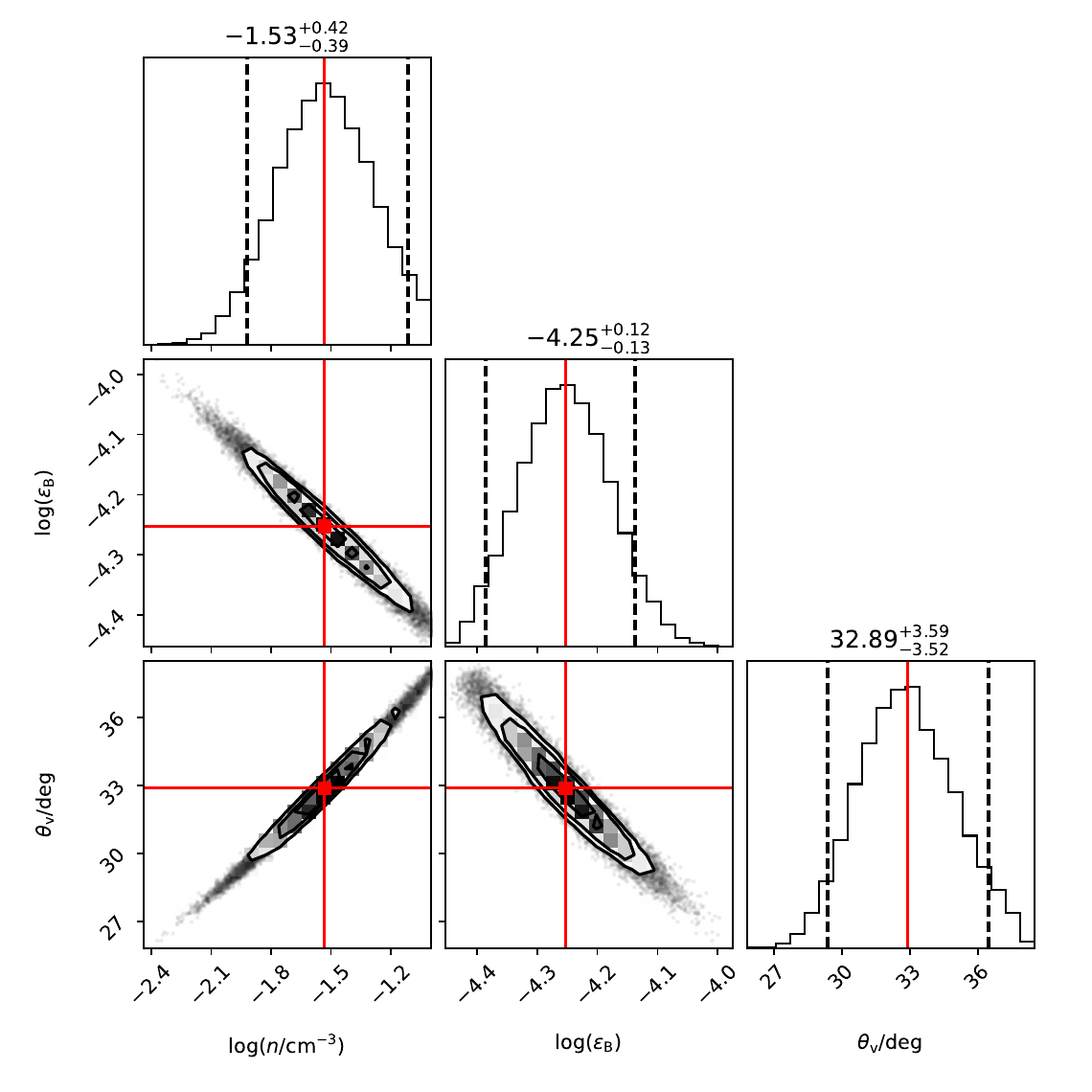}\includegraphics[width=0.5\textwidth]{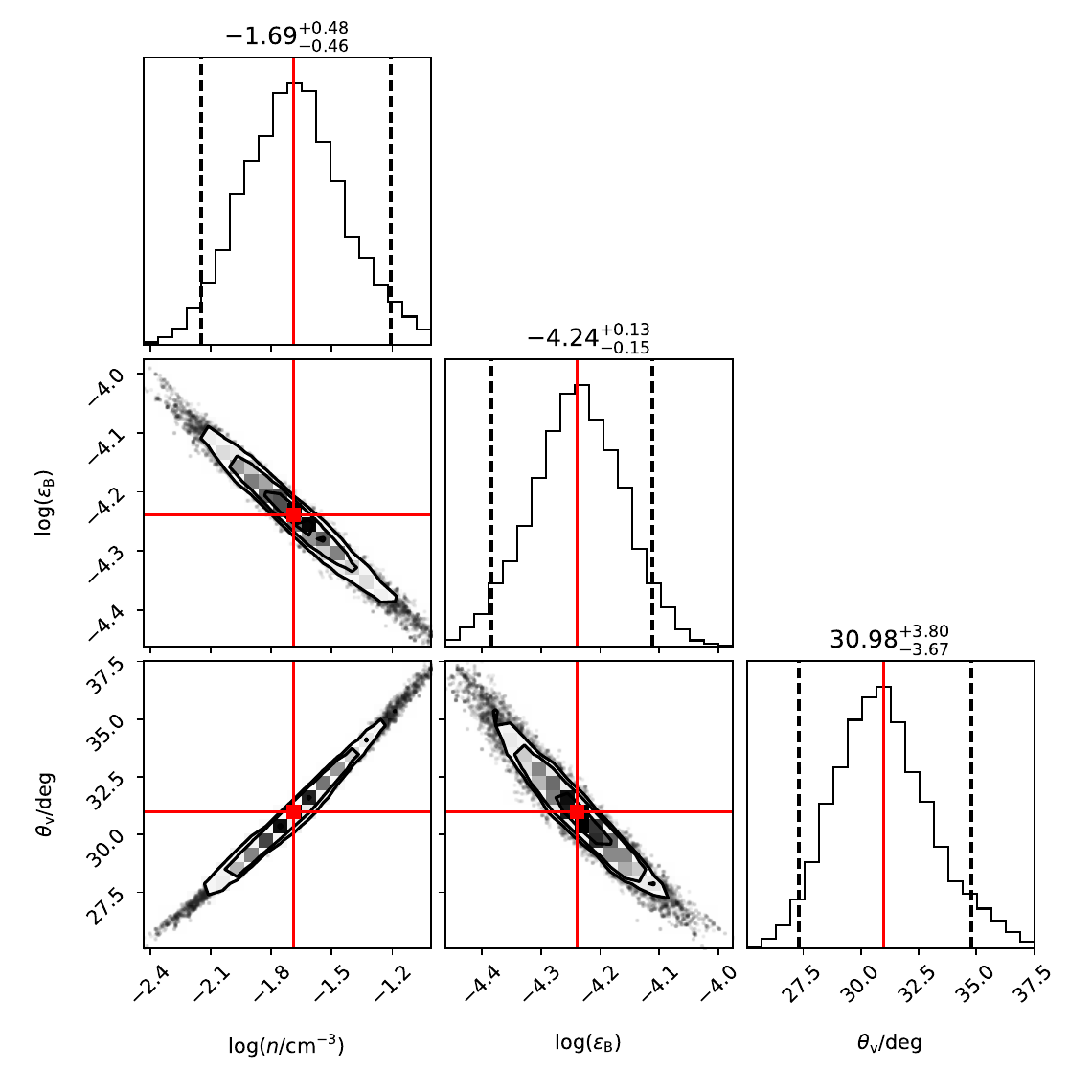}
    \caption{Corner plots of MCMC fit results of models $F$, $a$, and $b$ to the GW170817 data. Each panel shows the marginalized one-dimensional (histograms) and two-dimensional (contours) posterior probability distributions of the parameters of the MCMC fit of the afterglow from a given jet model ($F$: upper panel; $a$: lower left; $b$: lower right). Vertical black dashed lines show the 5$^\mathrm{th}$ and 95$^\mathrm{th}$ percentiles of the one-dimensional posteriors, while red solid lines show the positions of the medians. }
    \label{fig:corner_Fab}
\end{figure*}

\begin{figure*}
    \centering
    \includegraphics[width=0.5\textwidth]{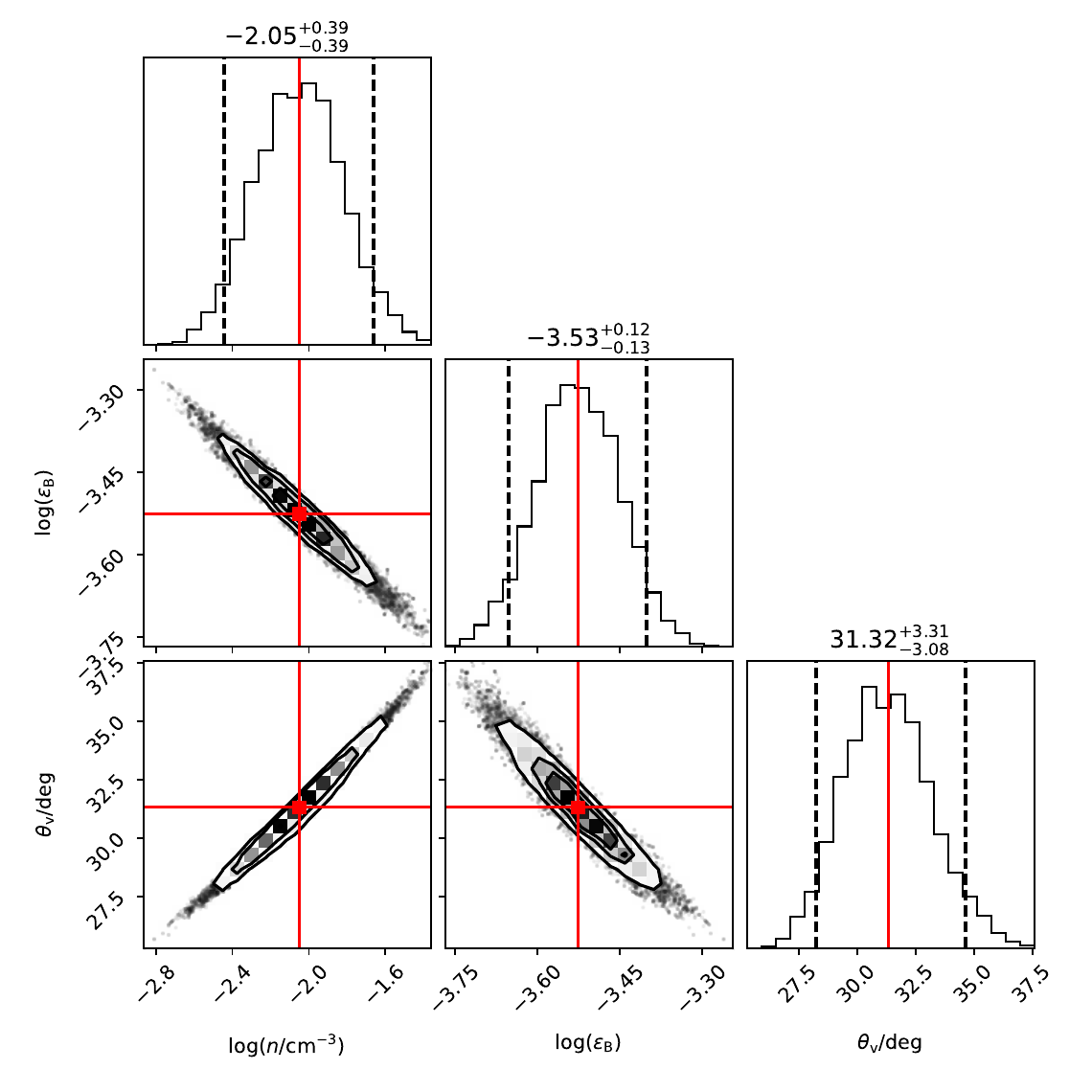}\includegraphics[width=0.5\textwidth]{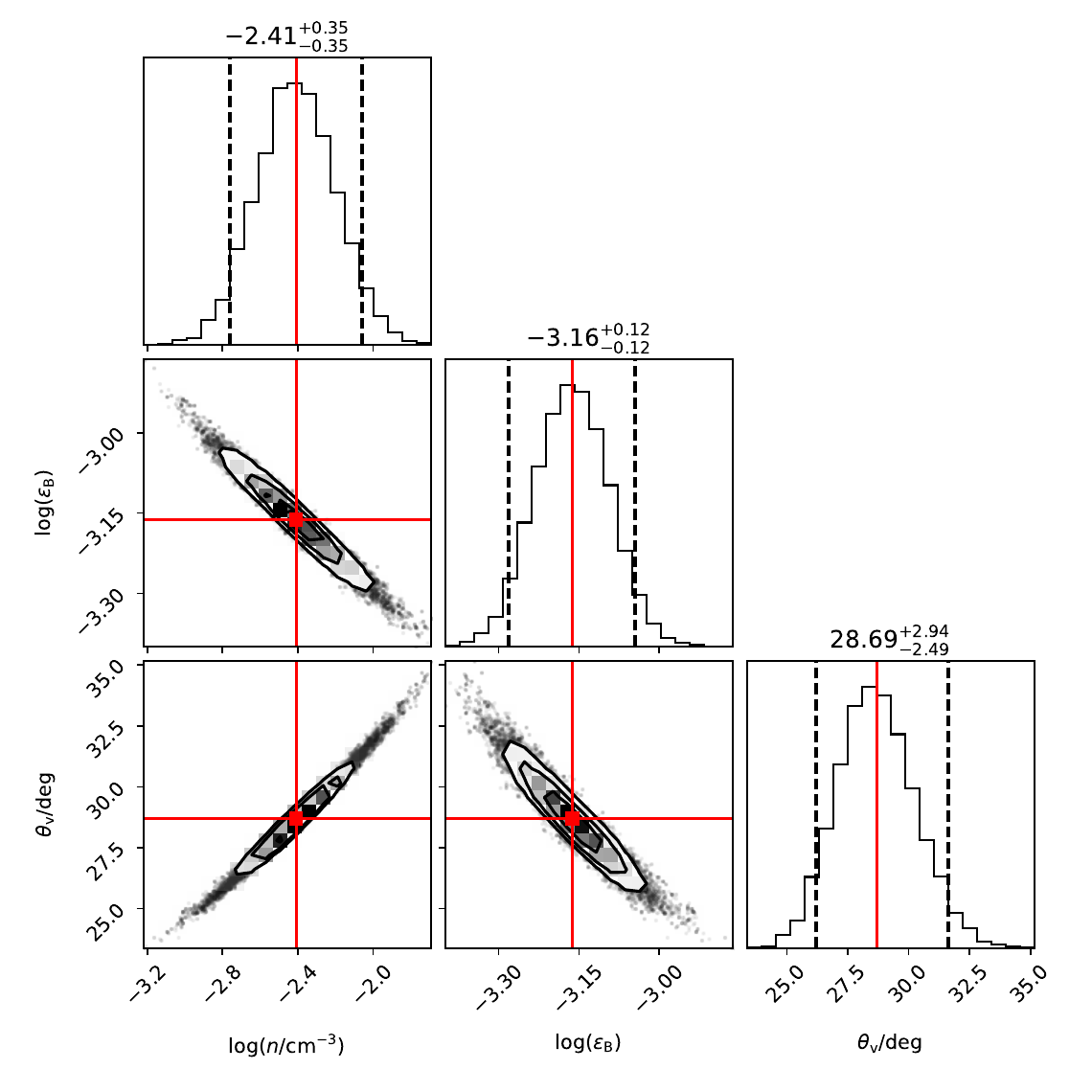}\\
    \includegraphics[width=0.5\textwidth]{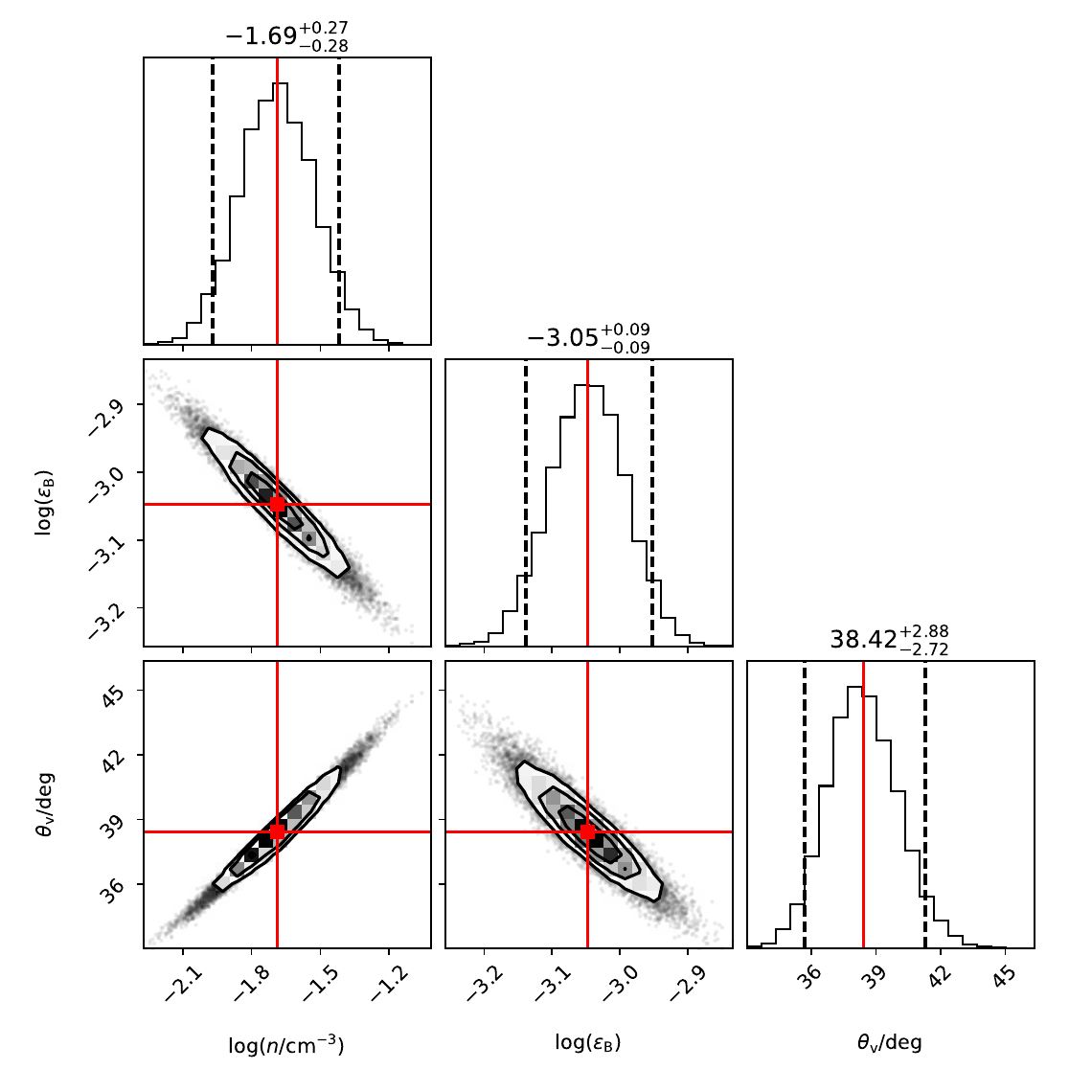}\includegraphics[width=0.5\textwidth]{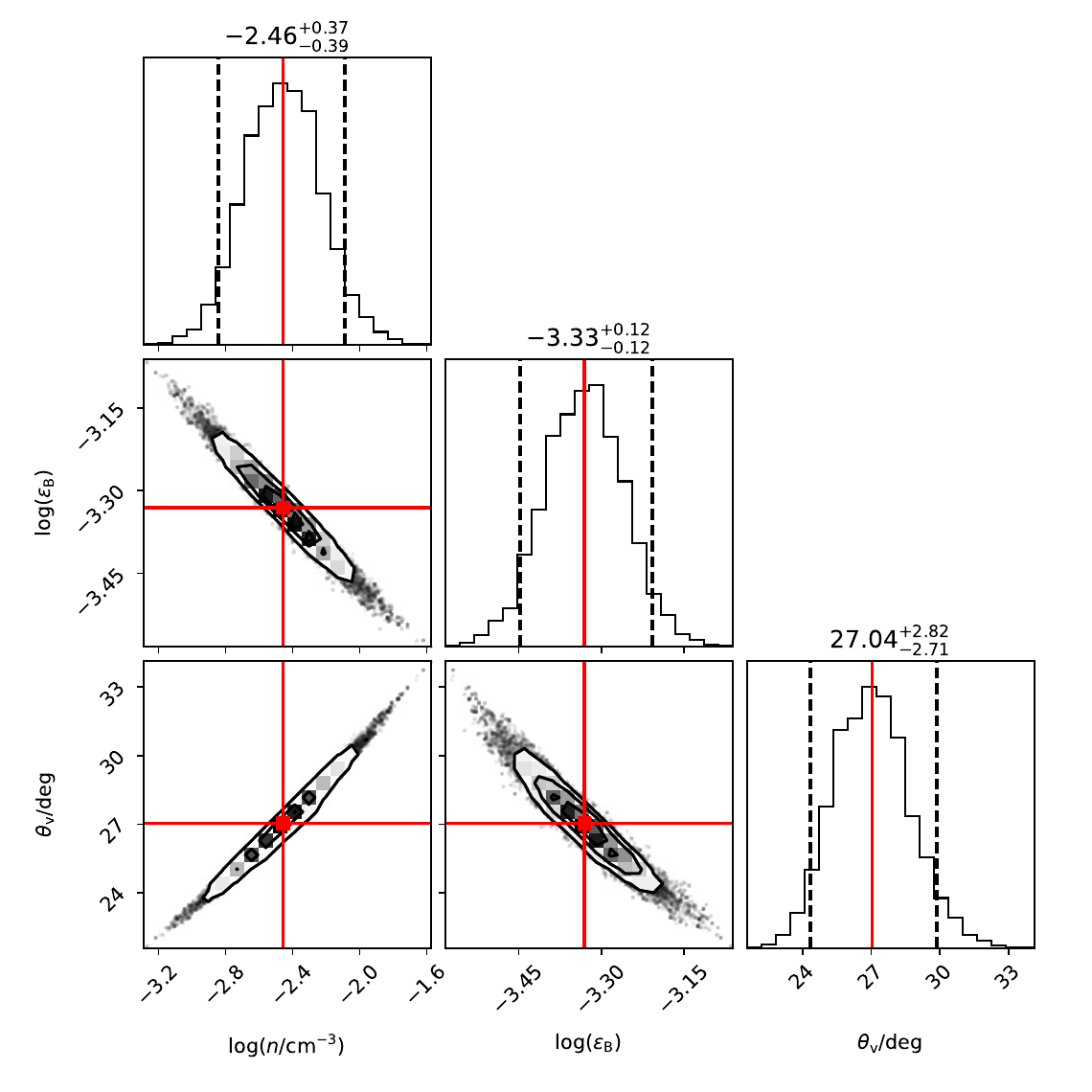}\\
    \caption{Corner plots of MCMC fit results of models $c$, $d$, $e1$ and $e2$ to the GW170817 data. Same as Fig.\ \ref{fig:corner_Fab}, but for models $c$ (upper left), $d$ (upper right), $e1$ (lower left) and $e2$ (lower right).}
    \label{fig:corner_cde1e2}
\end{figure*}

Figures \ref{fig:corner_Fab} and \ref{fig:corner_cde1e2} present the corner plots resulting from the MCMC fitting procedure to the data of GW170817 using the axisymmetrised jet structures of the different models, as described in Section~\ref{sec:fits}.

\begin{figure*}
    \centering
    \includegraphics[width=0.5\textwidth]{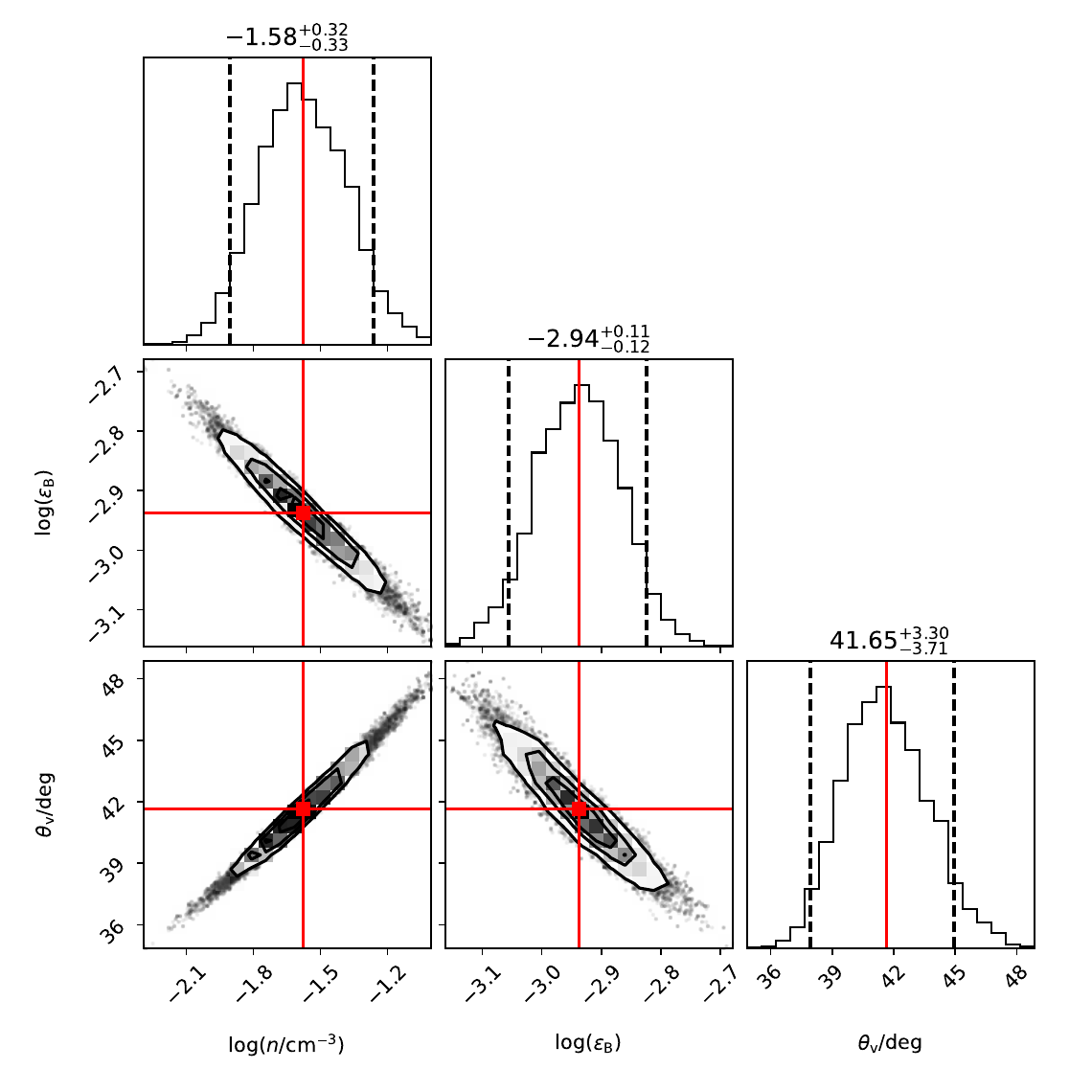}\includegraphics[width=0.5\textwidth]{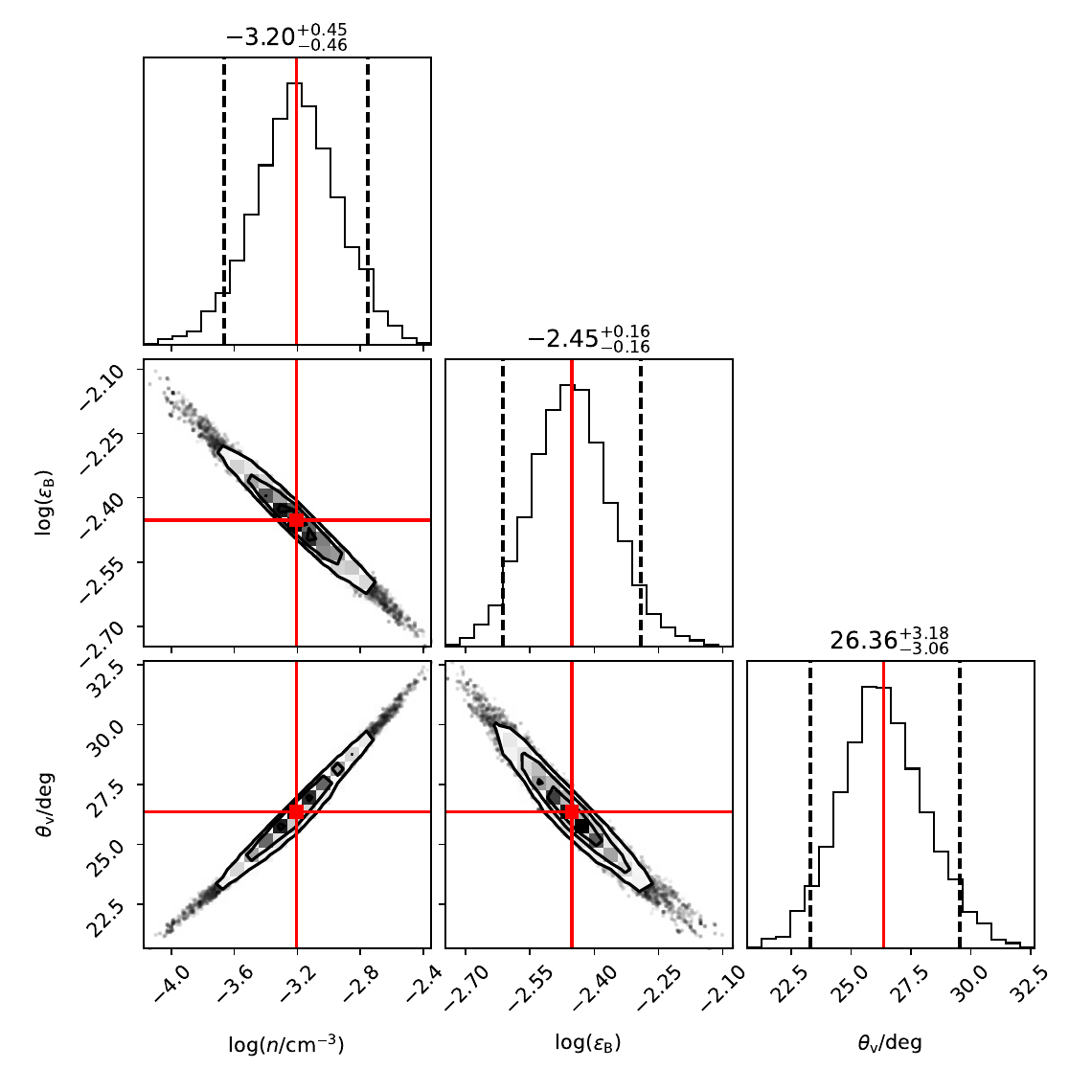}\\
    \includegraphics[width=0.5\textwidth]{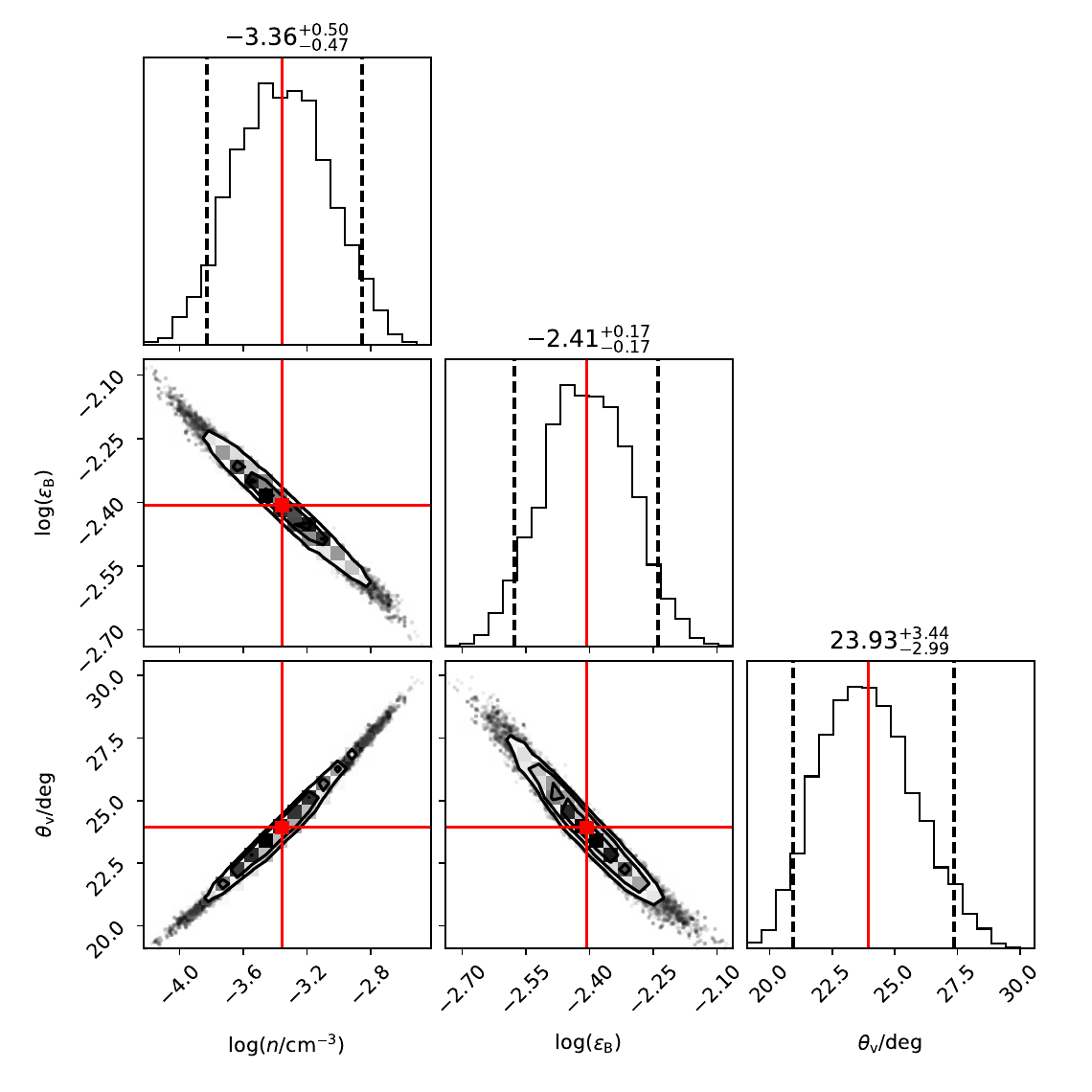}\\
    \caption{Corner plots of MCMC fit results of model $F$ to the GW170817 data. Same as Fig.\ \ref{fig:corner_Fab}, but considering the non-axisymmetrised jet structure and three different azimuthal viewing angles, i.e.~$\phi_v=0$ (upper left), $\phi_v=\pi /2$ (upper right), $\phi_v=\pi$ (bottom).}
    \label{fig:corner_Fview}
\end{figure*}

Figure \ref{fig:corner_Fview} presents the corner plots relative to the analogous MCMC fitting precedure using the non-axisymmetrised jet structure of model $F$ and considering three different azimuthal viewing angles, as described in Section ~\ref{sec:lc_asymmetry}.


\section{Resolution study}
\label{app:resolution}

Here we assess the dependence of our results on the numerical resolution of the jet simulations.

We focus first on model \textit{e1}, whose marginally successful breakout (see Section~\ref{sec:inj_time}) makes it a particularly sensitive test case for assessing resolution effects, and we compare it to model \textit{e2}, which only differs by the launching time. The simulation was re-run at a resolution increased by a factor of 1.2 along each spatial direction with respect to the fiducial setup adopted throughout this work (Section~\ref{sec:setup}). Table~\ref{tab:e1_hr} compares the jet cap properties of the two models at the fiducial resolution, from Table~2 of \citetalias{Pavan2025}, with those obtained at the increased resolution (models \textit{e1}$_{\rm HR}$ and \textit{e2}$_{\rm HR}$), following the same definitions of $E_{\rm iso,max}$, $\overline{\Gamma}_{\rm max}$, $E_{\rm sum}/E_{\rm inj}$, $E_{\rm kin}/E_{\rm sum}$, and $\Theta_{50}$.

At higher resolution, model \textit{e1} retains a larger fraction of its injected energy ($E_{\rm sum}/E_{\rm inj}$ increases from 0.44 to 0.47), converts a larger share of it into kinetic form ($E_{\rm kin}/E_{\rm sum}$ from 0.90 to 0.94), and reaches a substantially higher Lorentz factor ($\overline{\Gamma}_{\rm max}$ from 27.8 to 40.7), corresponding to a more successful, less marginal breakout. Consistently, the jet cap is somewhat more collimated at higher resolution ($\Theta_{50}$ from 6.6$^\circ$ to 5.9$^\circ$). 

\begin{figure*}
\centering
\includegraphics[width=1\textwidth]{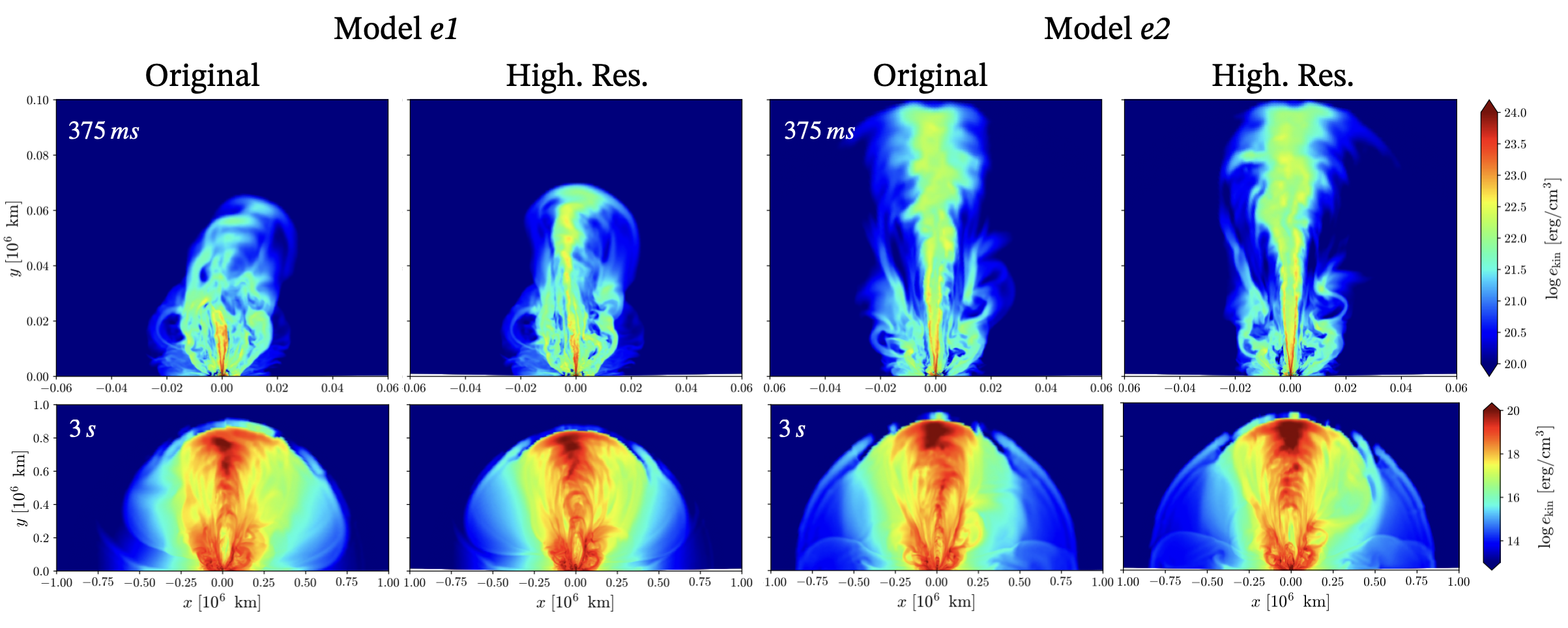}
\caption{Meridional view of $e_\mathrm{sum}$ at 375 ms (top) and 3 s (bottom) after jet launching, for model \textit{e1} (left) and \textit{e2} (right) at the two resolutions.}
\label{fig:e1_e2_hr}
\end{figure*}

Model \textit{e2} instead undergoes smaller modifications for all the parameters analysed, with variations at ten per cent level or below, most notably in the mean maximum Lorentz factor. The trend is nonetheless the same in both cases, the jet obtained at higher resolution being faster, more energetic and more collimated. 
Figure~\ref{fig:e1_e2_hr} compares meridional view of the energy density without rest mass component at an early stage in the evolution, corresponding to $t - t_{\rm jet} = 0.375\,\mathrm{s}$ and $t - t_{\rm jet} = 3\,\mathrm{s}$, for the two spatial resolutions considered in models \textit{e1} and \textit{e2}. At both times, visual inspection indicates that for model \textit{e1} the higher-resolution simulation confirms a more successful jet. For model \textit{e2} again, the differences are considerably less pronounced.

\begin{table}
\caption{Jet cap properties of models \textit{e1} and \textit{e2} at the fiducial resolution (from Table~2 of \citet{Pavan2025}, north side) and at increased resolution (\textit{e1}$_{\rm HR}$, \textit{e2}$_{\rm HR}$) at 3\,s.}
\label{tab:e1_hr}
\centering
\begin{tabular}{lcccc}
\hline
 & \textit{e1} & \textit{e1}$_{\rm HR}$ & \textit{e2} & \textit{e2}$_{\rm HR}$ \\
\hline
$E_{\rm iso,max}$ [$10^{53}$ erg] & 1.10 & 1.21 & 2.27 & 2.58 \\
$\overline{\Gamma}_{\rm max}$ & 27.8 & 40.7 & 70.6 & 76.2 \\
$E_{\rm sum}/E_{\rm inj}$ & 0.44 & 0.47 & 0.61 & 0.63 \\
$E_{\rm kin}/E_{\rm sum}$ & 0.90 & 0.94 & 0.92 & 0.92 \\
$\Theta_{50}$ [$^\circ$] & 6.58 & 5.89 & 4.41 & 4.06 \\
\hline
\end{tabular}
\end{table}

\begin{figure}
    \centering
    \includegraphics[width=1\linewidth]{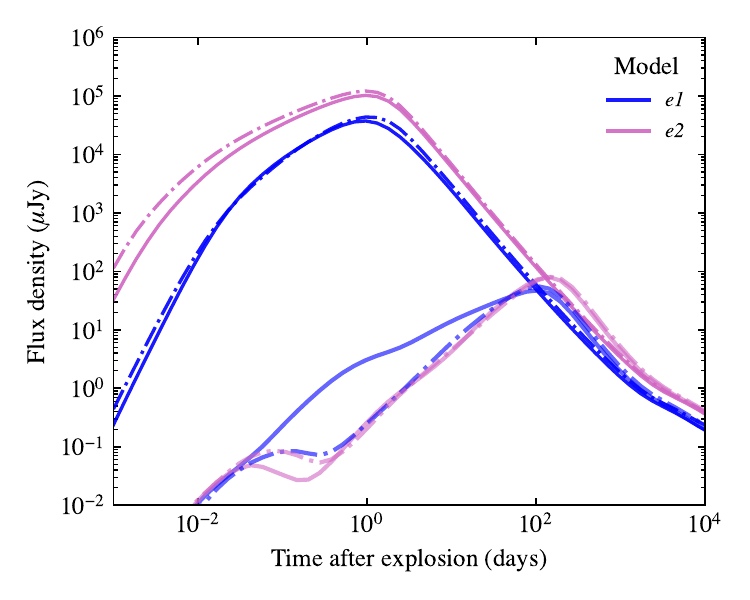}
    \caption{Afterglow light curves at 3\,GHz for model \textit{e1} (blue), \textit{e2} (pink) at the fiducial resolution (solid line) and increased resolution (dashed line), on-axis and with $\theta_\mathrm{v}=32^\circ$, computed with the same afterglow parameters (best-fit values of model \textit{F}) for both resolutions.}
    \label{fig:lc_hr}
\end{figure}

The corresponding on-axis light curves, shown in Fig.~\ref{fig:lc_hr}, demonstrate that, despite the non-negligible structural modification, the resulting afterglow signal retains the same qualitative shape across resolutions, indicating that near-axis predictions are comparatively robust to numerical resolution. Off-axis, however, the higher-resolution run shows a qualitative change for model \textit{e1}: the extended pre-peak plateau that, at fiducial resolution, sets it apart from the other, more energetic models becomes markedly less pronounced, with the rising shape approaching that of the other models. However, the location of the peak remains mostly unchanged, and its flux slightly increased. The curve of model \textit{e2} remains instead mostly unchanged, consistently with the previous analysis. 
For the two models considered here, the change in resolution modifies the shape of the light curves, while the relative difference between the two models is retained. We expect a similar behaviour for the other models, for which we do not repeat the analysis. This implies that differences reported between models, including some of a qualitative nature, that are comparable to those arising from numerical resolution cannot be firmly disentangled from resolution effects.

\section{Detectability of the outer-shell feature}
\label{sec:shell_detectability}

In this appendix we discuss the detectability of the observational feature associated with the outer shell, introduced in Sect.~\ref{sec:2shell} and further discussed in Sect.~\ref{sec:lc_velprof} in the context of the afterglow light curves.

Drawing a firm conclusion on this point is not straightforward, since the flux density of the early peak depends on several parameters at once. The source distance has the strongest influence, but the flux also depends on the ISM density and on the observing angle, both in the axisymmetrised case and when different azimuthal viewing angles are considered. Different jet structures lead to considerable differences in this feature as well (see Figs.~\ref{fig:lc_velprof} and \ref{fig:lc_sym_OFF}).

\begin{figure}
    \centering
    \includegraphics[width=1\linewidth]{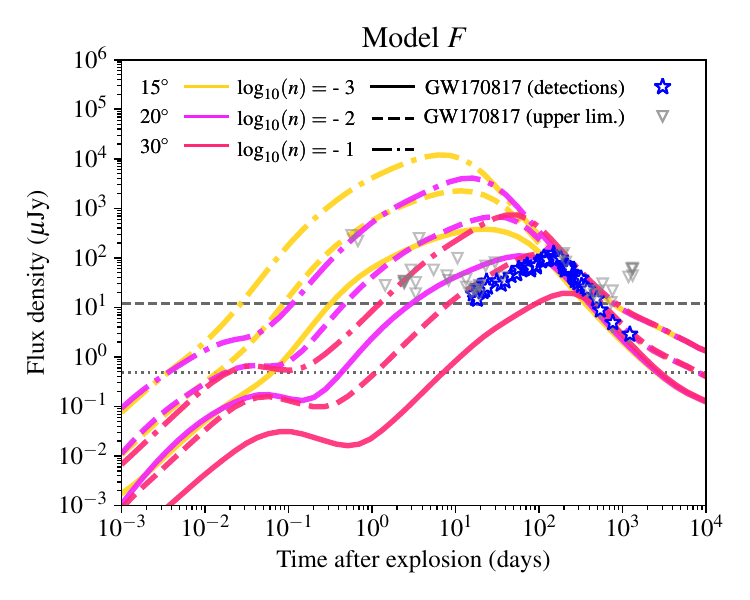}
    \caption{Off-axis afterglow light curves at 3~GHz for model \textit{F} (axisymmetrised jet structure), for a set of viewing angles $\theta_v = 15^\circ, 20^\circ, 30^\circ$ (colours) and ISM densities $n = 10^{-3}, 10^{-2}, 10^{-1}~\mathrm{cm^{-3}}$ (line styles), illustrating the dependence of the early shell-associated peak on these parameters. Horizontal dashed and dotted grey lines mark the $3\sigma$ VLA and ngVLA sensitivities, respectively, for a fiducial 1~hr integration at 3~GHz. Grey stars with error bars and downward-pointing arrows show, respectively, GW170817 radio detections and $3\sigma$ upper limits at frequencies between 1 and 10~GHz, rescaled to 3~GHz, as compiled by \citet{Makhathini2021} from \citet{Alexander2017} and \citet{Hallinan2017}.}
    \label{fig:shell_detectability}
\end{figure}

Focusing on the radio band, we compare the $3\sigma$ VLA \citep{Perley2011} and ngVLA \citep{Selina2018} sensitivities to the off-axis light curves obtained from the axisymmetrised structure of model \textit{F}, for a set of viewing angles and ISM densities (Fig.~\ref{fig:shell_detectability}). The $3\sigma$ VLA and ngVLA sensitivity thresholds are computed for a fiducial $1$~h on-source integration at $3$~GHz using the respective official exposure calculators, 
adopting dual polarisation and natural weighting. For the VLA in S-band 
($2$--$4$~GHz) this yields a $1\sigma$ rms of $\simeq 4~\mu\mathrm{Jy}/\mathrm{beam}$, 
and for the ngVLA reference design in Band~1 (main+long subarray, full 
$2.30$~GHz bandwidth) $\simeq 0.16~\mu\mathrm{Jy}/\mathrm{beam}$, an 
order-of-magnitude improvement. We also show the early GW170817 upper limits and non-detections, taken from \citet{Alexander2017} and \citet{Hallinan2017} as compiled in the uniform dataset of \citet{Makhathini2021}. We find that, for the highest densities considered, combined with an intermediate viewing angle, the peak of the feature crosses the ngVLA detection threshold. In general, however, it remains well below the level of the early-time GW170817 upper limits, indicating that these do not yet provide a meaningful constraint on this specific signature.

A further layer of complexity arises from azimuthal asymmetries and from differences among the various jet models. To give a qualitative sense of this additional variability, we overlay the same sensitivity lines on the off-axis light curves of all models, including the spread introduced by different azimuthal viewing angles, in Fig.~\ref{fig:symm_all}.
While some of these curves appear promising for future facilities, an additional, more fundamental constraint is set by time. For model \textit{F}, the outer-shell peak occurs at $t\sim10^{-2}$--$10^{-1}$~d after the merger (i.e.\ between roughly a quarter of an hour and two and a half hours). This is representative of the general behaviour: across all models the feature peaks at comparably early times, although the precise value varies considerably from model to model, reflecting the dependence on the jet injection parameters discussed above. In every case, however, this timescale is comparable to, or shorter than, the 1~hr integration adopted here to build the sensitivity curves. Even the most rapid radio follow-up observations obtained to date for a BNS merger remain insufficient to meet this requirement: the electromagnetic counterpart of GW170817 was itself identified remarkably early, within $\sim$11~h of the merger \citep{Coulter2017}, and the earliest available VLA observation followed at $t\simeq0.57$~d \citep{Alexander2017}, only a few hours later. This response, on a timescale of hours rather than days, was exceptionally fast by the standards of the time; yet it still follows after the predicted peak of model \textit{F} by roughly eleven hours. Reaching such early timescales would require sub-hour localisation-to-pointing response, and radio follow-up capabilities have continued to improve in the past years with faster alert communication, progressively narrowing this gap. Moreover, even under such favourable circumstances, if the intrinsic duration of the feature is comparable to or shorter than the required integration time, the resulting flux estimate would effectively average over a rapidly evolving source, further diluting the sensitivity to this specific signature. The detectability of the outer-shell feature is therefore limited jointly by flux threshold and by timing. It nonetheless remains a potential diagnostic for structured jets, that could become more readily accessible as higher-sensitivity facilities, faster and more precise localisation, and closer events become available.

\end{document}